\documentclass[fleqn,usenatbib]{mnras}

\usepackage{newtxtext,newtxmath}

\usepackage[T1]{fontenc}

\DeclareRobustCommand{\VAN}[3]{#2}
\let\VANthebibliography\thebibliography
\def\thebibliography{\DeclareRobustCommand{\VAN}[3]{##3}\VANthebibliography}

\usepackage{amsmath}
\usepackage{microtype}
\usepackage{booktabs}
\usepackage{graphicx}	
\usepackage{color}
\usepackage{arydshln}
\usepackage{natbib}
\usepackage{xspace}
\usepackage{xcolor}
\usepackage{hyperref}
\hypersetup{
    colorlinks=true,
    linkcolor=blue,
    citecolor=blue,
    filecolor=blue,      
    urlcolor=blue,
    pdftitle={}
    }

\usepackage{multirow}

\newcommand{\Msun}{\ensuremath{\mathrm{M}_{\odot}}\xspace}
\newcommand{\Msunpc}{\ensuremath{\Msun\, \mathrm{pc}^{-2}}\xspace}
\newcommand{\Msunyr}{\ensuremath{\Msun\, \mathrm{yr}^{-1}}\xspace}
\newcommand{\kms}{\ensuremath{\text{km} \, \text{s}^{-1}}\xspace}

\newcommand{\SN}{\ensuremath{S/N}\xspace}

\newcommand{\HI}{{H}\,{\sc i}\xspace}

\newcommand{\HII}{{H}\,{\sc ii}\xspace}
\newcommand{\HA}{\ensuremath{\mathrm{H}{\alpha}}\xspace}
\newcommand{\HB}{\ensuremath{\mathrm{H}{\beta}}\xspace}
\newcommand{\OIII}{\ensuremath{\mathrm{[O\textsc{iii}]}}\xspace}
\newcommand{\OIIIb}{\ensuremath{\mathrm{[O\textsc{iii}]_{4959}}}\xspace}
\newcommand{\OIIIr}{\ensuremath{\mathrm{[O\textsc{iii}]_{5007}}}\xspace}
\newcommand{\NII}{\ensuremath{\mathrm{[N\textsc{ii}]}}\xspace}
\newcommand{\NIIb}{\ensuremath{\mathrm{[N\textsc{ii}]_{6548}}}\xspace}
\newcommand{\NIIr}{\ensuremath{\mathrm{[N\textsc{ii}]_{6583}}}\xspace}
\newcommand{\SII}{\ensuremath{\mathrm{[S\textsc{ii}]}}\xspace}
\newcommand{\SIIb}{\ensuremath{\mathrm{[S\textsc{ii}]_{6716}}}\xspace}
\newcommand{\SIIr}{\ensuremath{\mathrm{[S\textsc{ii}]_{6731}}}\xspace}

\newcommand{\logOH}{\ensuremath{\log(O/H)}\xspace}
\newcommand{\TWlogOH}{\ensuremath{12+\log(O/H)}\xspace}
\newcommand{\ergscm}{\ensuremath{{\mathrm{erg} \,\mathrm{s}^{-1} \,\mathrm{cm}^{-2}}}\xspace}

\newcommand{\lgMstarMsun}{\ensuremath{\log(M_{\star}/\Msun)}\xspace}

\newcommand{\lgMHI}{\ensuremath{\log(M_{\text{\HI}}/\Msun)}\xspace}

\newcommand{\lgSFRMsunyr}{\ensuremath{\log\,\mathrm{SFR}\,[\Msunyr]}\xspace}

\newcommand{\RHI}{\ensuremath{R_\text{\HI}}\xspace}

\newcommand{\SigSFR}{\ensuremath{\Sigma_\mathrm{SFR}}\xspace}
\newcommand{\lgSigSFR}{\ensuremath{\log\Sigma_\mathrm{SFR}}\xspace}

\newcommand{\lgSigHIMsun}{\ensuremath{\log(\Sigma_\text{\HI}/\Msunpc)}\xspace}

\newcommand{\cp}{\citep}
\newcommand{\ct}{\citet}
\newcommand{\fig}[1]{Fig.~\ref{fig:#1}}
\newcommand{\ts}[1]{\textsuperscript{#1}}

\title[MAUVE: The ionised gas outflow of NGC4383]{MAUVE: A 6\,kpc bipolar outflow launched from NGC\,4383, one of the most \HI-rich galaxies in the Virgo cluster}

\author[A.B. Watts et al.]{
Adam B. Watts$^{1,2}$,\thanks{E-mail: adam.watts@uwa.edu.au}
Luca Cortese$^{1,2}$,
Barbara Catinella$^{1,2}$,
Amelia Fraser-McKelvie$^{1,2}$,
Eric Emsellem$^{3,4}$,
\newauthor{
Lodovico Coccato$^{3}$,
Jesse van de Sande$^{2,5}$,
Toby H. Brown$^{6}$,
Yago Ascasibar$^{7}$,
Andrew Battisti}$^{2,8}$,
\newauthor{
Alessandro Boselli$^{9}$,
Timothy A. Davis$^{10}$,
Brent Groves$^{1,2}$,
and Sabine Thater$^{11}$}
\\
$^{1}$International Centre for Radio Astronomy Research, The University of Western Australia, Crawley, WA 6009, Australia\\
$^{2}$ARC Centre of Excellence for All-Sky Astrophysics in 3 Dimensions (ASTRO3D), Australia\\
$^{3}$European Southern Observatory, Karl-Schwarzschild-Straße 2, 85748 Garching, Germany \\
$^{4}$Univ Lyon, Univ Lyon1, ENS de Lyon, CNRS, Centre de Recherche Astrophysique de Lyon UMR5574, F-69230 Saint-Genis-Laval France\\
$^{5}$Sydney Institute for Astronomy (SIfA), School of Physics, The University of Sydney, NSW 2006, Australia\\
$^{6}$Herzberg Astronomy and Astrophysics Research Centre, National Research Council of Canada, 5071 West Saanich Rd., Victoria, BC V9E 2E7, Canada\\
$^{7}$Departamento de Física Teórica, Universidad Autónoma de Madrid (UAM), Campus de Cantoblanco, E-28049 Madrid, Spain\\
$^{8}$Research School of Astronomy and Astrophysics, Australian National University, Canberra, ACT 2611, Australia\\
$^{9}$Aix-Marseille Universit\'{e}, CNRS, CNES, LAM, Marseille, France\\
$^{10}$Cardiff Hub for Astrophysics Research \& Technology, School of Physics \& Astronomy, Cardiff University, Queens Buildings, Cardiff, CF24 3AA, UK\\
$^{11}$Department of Astrophysics, University of Vienna, Türkenschanzstraße 17, 1180, Vienna, Austria
}

\date{Accepted XXX. Received YYY; in original form ZZZ}

\pubyear{2023}

\begin{document}
\label{firstpage}
\pagerange{\pageref{firstpage}--\pageref{lastpage}}
\maketitle

\begin{abstract}
Stellar feedback-driven outflows are important regulators of the gas-star formation cycle.
However, resolving outflow physics requires high resolution observations that can only be achieved in very nearby galaxies, making suitable targets rare. 
We present the first results from the new VLT/MUSE large program MAUVE (MUSE and ALMA Unveiling the Virgo Environment), which aims to understand the gas-star formation cycle within the context of the Virgo cluster environment.
Outflows are a key part of this cycle, and we focus on the peculiar galaxy NGC\,4383, which hosts a $\sim6\,$kpc bipolar outflow fuelled by one of Virgo's most \HI-rich discs. 
The spectacular MUSE data reveal the clumpy structure and complex kinematics of the ionised gas in this M82-like outflow at 100\,pc resolution.
Using the ionised gas geometry and kinematics we constrain the opening half-angle to $\theta=25-35^\circ$, while the average outflow velocity is $\sim210\,\kms$. 
The emission line ratios reveal an ionisation structure where photoionisation is the dominant excitation process. 
The outflowing gas shows a marginally elevated gas-phase oxygen abundance compared to the disc but is lower than the central starburst, highlighting the contribution of mixing between the ejected and entrained gas. 
Making some assumptions about the outflow geometry, we estimate an integrated mass outflow-rate of $\sim1.8~\Msunyr$ and a corresponding mass-loading factor in the range 1.7-2.3. 
NGC\,4383 is a useful addition to the few nearby examples of well-resolved outflows, and will provide a useful baseline for quantifying the role of outflows within the Virgo cluster. 
\end{abstract}




\section{Introduction}
Star formation in galaxies proceeds under a self-regulated balance of accretion, consumption, and ejection of gas between the interstellar medium (ISM), and the circumgalactic medium (CGM) and intergalactic medium \cp{sancisi08,lilly13,fraternali17,tumlinson17,wright21,saintonge22}.
Feedback from star formation is a key physical mechanism for this self regulation \cp[e.g.,][]{krumholz18,mcleod21,ostriker22,egorov23}, and the signatures of star formation-driven outflows are ubiquitous in star-forming galaxies \cp[e.g.][]{chen10,rubin14,arribas14}.
Driven by the winds of high mass stars and supernovae explosions, stellar feedback is responsible for dissociating molecular clouds and halting star formation locally \cp{jeffreson21,chevance20a,barnes23}, and distributing heavy elements throughout galaxies. 
In more extreme cases of galaxies with intense central bursts of star formation, the resulting feedback can drive large outflows of metal-enriched gas from the centres of galaxies far into their CGM \cp[e.g.,][]{heckman00,tremonti04,borthakur13,chisholm18,peroux20}.
Understanding the physics of stellar feedback-driven outflows, thus, is essential to completing our picture of galaxy evolution. 

Stellar feedback-driven outflows are commonly thought to be generated by the spatial and temporal clustering of supernovae, which each injects several $\Msun$ of material with $E~\sim10^{51}$\,erg into the surrounding ISM \cp[e.g.][]{chevalier85,nguyen23}. 
These supernovae drive expanding bubbles of hot gas, which break out from the disc vertically due to the stratified structure of the ISM in disc galaxies \cp[e.g.,][]{cooper08,fielding17,fielding18,orr22}.
During its expansion this hot gas sweeps up and entrains the cooler phases of the ISM, increasing the mass of ejected gas (mass-loading) and carrying it into the CGM \cp[e.g.,][]{thompson16,gronke18,gronke20,schneider20}. 
Thus, outflows are inherently multiphase \cp[see the reviews by][]{rupke18,veilleux20}, where the majority of the outflow energy is contained within a hot ($T>10^6$~K) volume filling phase, while the mass is carried by colder ($T\lesssim10^4$~K) neutral atomic and molecular gas entrained in the outflow \cp{leroy15,martin-fernandez16,kim18, yuan23}.
Subsequently, outflows are observed in emission in X-rays \cp[e.g.,][]{lehnert99,hodges-kluck20} and cold gas tracers like CO, neutral atomic hydrogen (\HI), and dust \cp[e.g.,][]{seaquist01,melendez15, leroy15,martini18,stuber21}, while the mixing and cooling of these phases gives rise to nebular emission lines through the ultraviolet, optical, and infrared \cp[e.g.][]{veilleux94,fogarty12,ho16a,cicone18,davies19,rupke23}.
Gas ejected by outflows is also detected in absorption against bright background UV and optical sources \cp[e.g.,][]{hota05,weiner09,erb12,avery22}, enabling their easier study to higher redshift and galactocentric radii. 

Despite their multiwavelength signatures, the physics of outflows has remained hard to constrain. 
Absorption line studies have been essential for establishing and testing the dependence (or lack thereof) of the gas velocity and mass loading factor (the rate of outflowing gas mass normalised by the current star-formation rate, SFR: $\eta_M=\dot{M}_\mathrm{out}/\mathrm{SFR}$) on host galaxy properties such as stellar mass, SFR, and circular velocity \cp[e.g.,][]{martin05a,kornei12,chisholm15,heckman16,xu22a}.
However, they contain little to no spatially resolved information, which carries the signatures of the details of outflow physics. 
The morphology and multiphase structure of the gas within outflows is sensitive to the ISM conditions around their launching sites \cp[e.g.,][]{creasey13,girichidis16,fielding18}, the details of mixing and ISM entrainment \cp[e.g.,][]{ji19,hu19,schneider20,fielding22},  the influence of magnetic fields and cosmic rays \cp[e.g.,][]{sparre20,lopez-rodriguez21,armillotta22,peschken23,rathjen23}.

Spatially resolving outflows however, presents its own challenges. 
Their diffuse nature means that they are much fainter than their host galaxies, requiring many hours on the most sensitive telescopes to detect and characterise. 
Further, observing sufficiently small scales to study the interactions between their multiphase components requires high resolution observations of nearby objects, making suitable systems rare. 
For these reasons, our observational understanding of the spatially resolved nature of outflows has been limited to a few nearby systems. 
M\,82 is the archetypal example of a central starburst-driven outflow, and subsequently has been the subject of numerous multiwavelength studies \cp[e.g.,][]{shopbell98,walter02a,veilleux09,leroy15,martini18,yuan23}. 
Other nearby systems include the narrow outflow from NGC\,253 \cp[e.g.,][]{ulrich78,bolatto13,walter17,krieger19,lopez23}, the hourglass-shaped and collisionally excited wind in the early-type starburst NGC\,1482 \cp[e.g.,][]{veilleux02, hota05,vagshette12,salak20} and several nearby dwarf galaxies \cp[e.g.,][]{meurer92,walter02,mcquinn19,egorov21,marasco23}; while slightly more distant examples are beginning to be resolved \cp[e.g.,][]{rich10,cameron21,reichardtchu22,reichardtchu22a,mcpherson23}.
Regardless, every new spatially resolved-outflow added to the list adds significant new information for improving our understanding of outflow physics.

In this paper, we present the first results of the MUSE and ALMA Unveiling the Virgo Environment (MAUVE) survey, a new ESO large program targeting Virgo cluster galaxies with the Multi Unit Spectroscopic Explorer \cp[MUSE,][]{bacon10} on the VLT.
Combined with \HI data from the VIVA\footnote{VLA Imaging of Virgo spirals in Atomic gas.} survey \cp{chung09}, molecular gas from the VERTICO\footnote{Virgo Environment Traced in CO.} survey \cp{brown21}, and the rich multiwavelength coverage afforded by Virgo being our nearest galaxy cluster \cp[e.g.,][]{york00,davies10,urban11,boselli11,boselli18a}, MAUVE will provide a comprehensive view of the physical nature of star formation quenching in galaxy clusters. 
Our analysis focuses on the MUSE observations of a pilot program target, NGC\,4383, which hosts a massive bipolar outflow of ionised gas.
We describe our target in the next section (\S\ref{subsec:history}), and our observations, data reduction, and measurement of the data products in \S\ref{sec:DR}. 
In \S\ref{sec:outflow} we present the outflow geometry and kinematics, estimate the metallicity of the outflow, and the mass of outflowing gas, before placing NGC\,4383 in context with the literature and concluding in \S\ref{ref:concl}. 
Throughout this work we assume that NGC\,4383 is at the distance of Virgo, 16.5\,Mpc \cp{mei07}, which corresponds to a spatial resolution of 80\,pc per arcsecond.

\subsection{NGC\,4383} \label{subsec:history}
\begin{figure*}
    \includegraphics[width=0.48\textwidth]{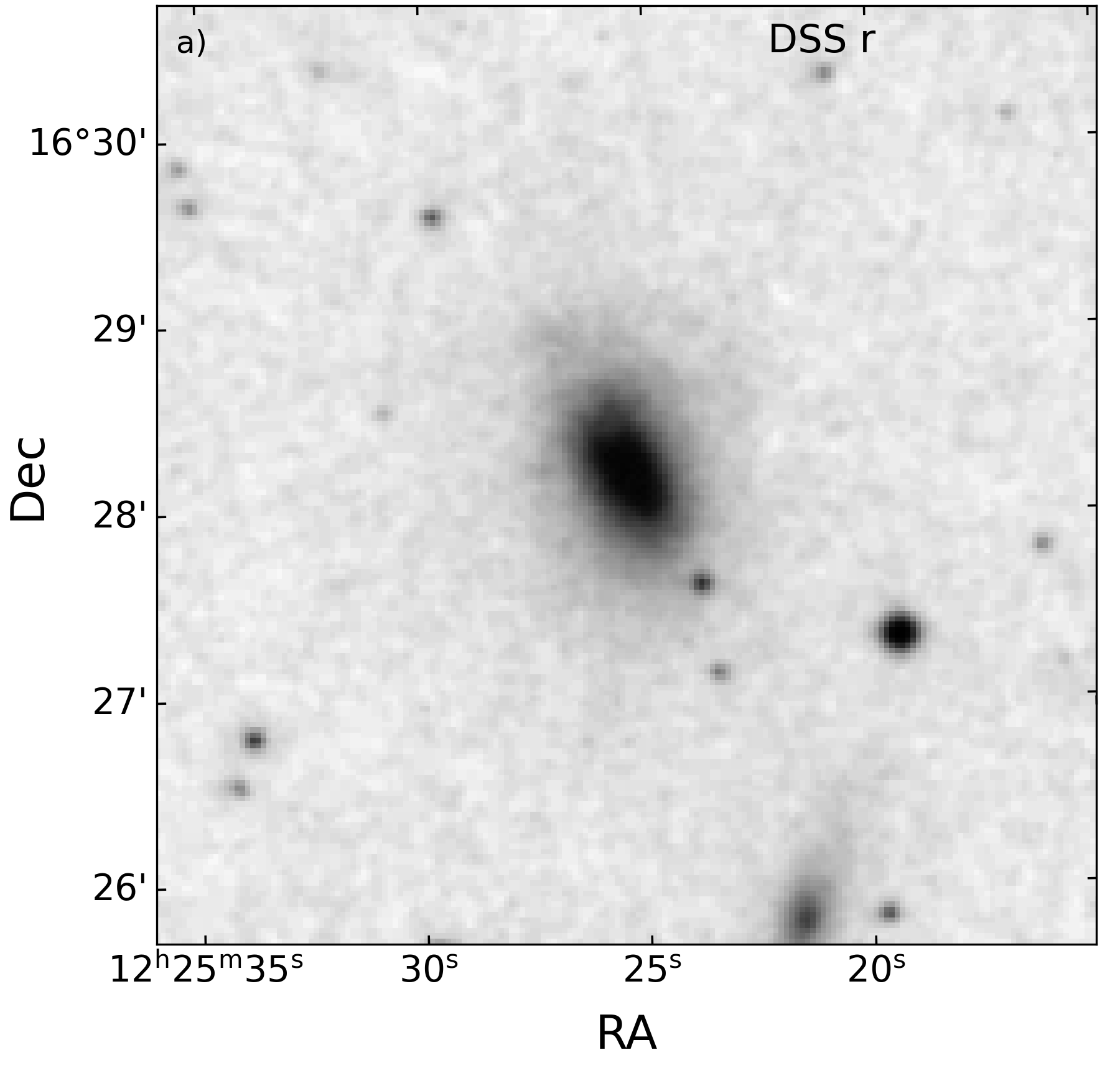}
    \includegraphics[width=0.48\textwidth]{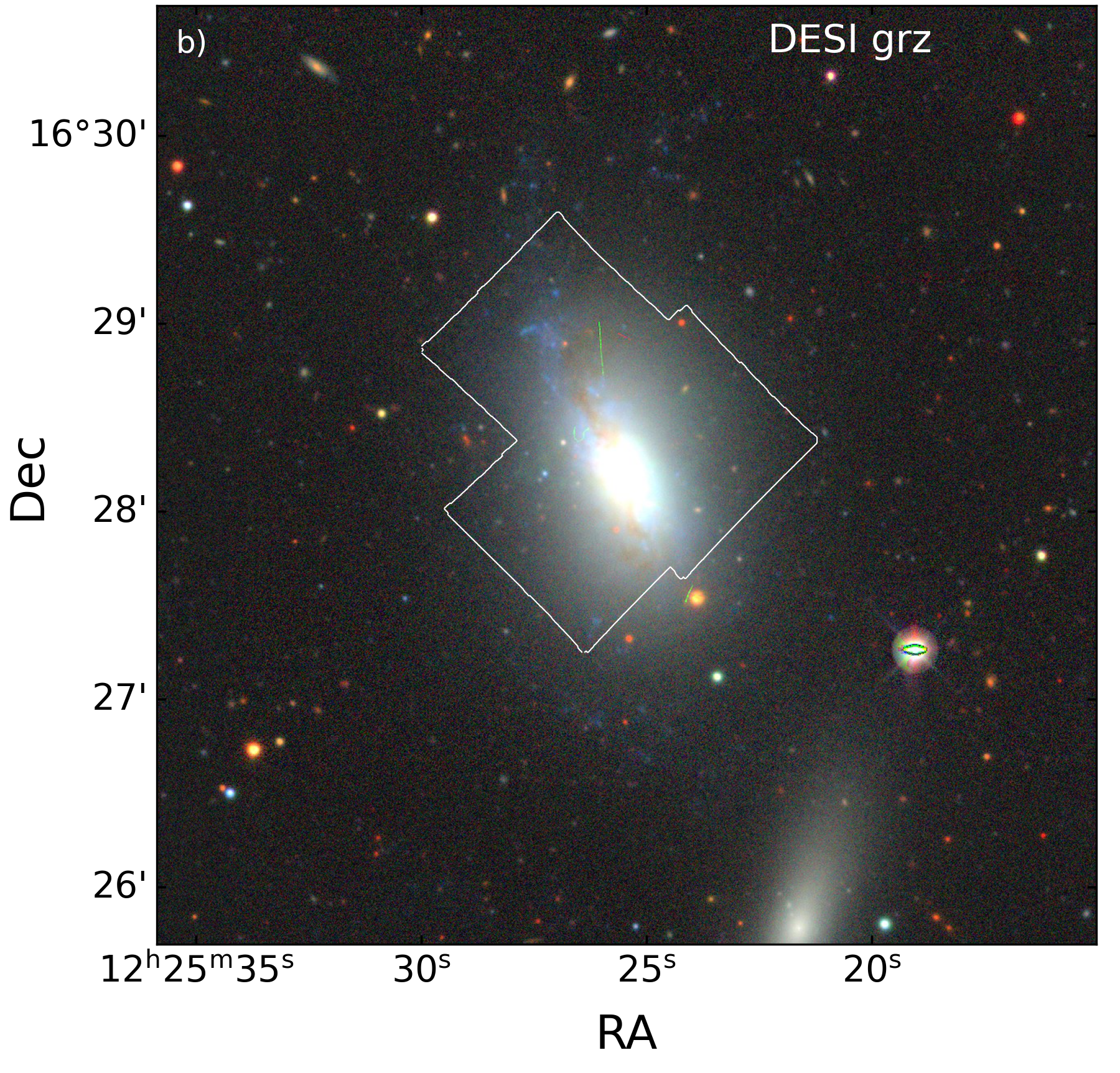}
    \includegraphics[width=0.48\textwidth]{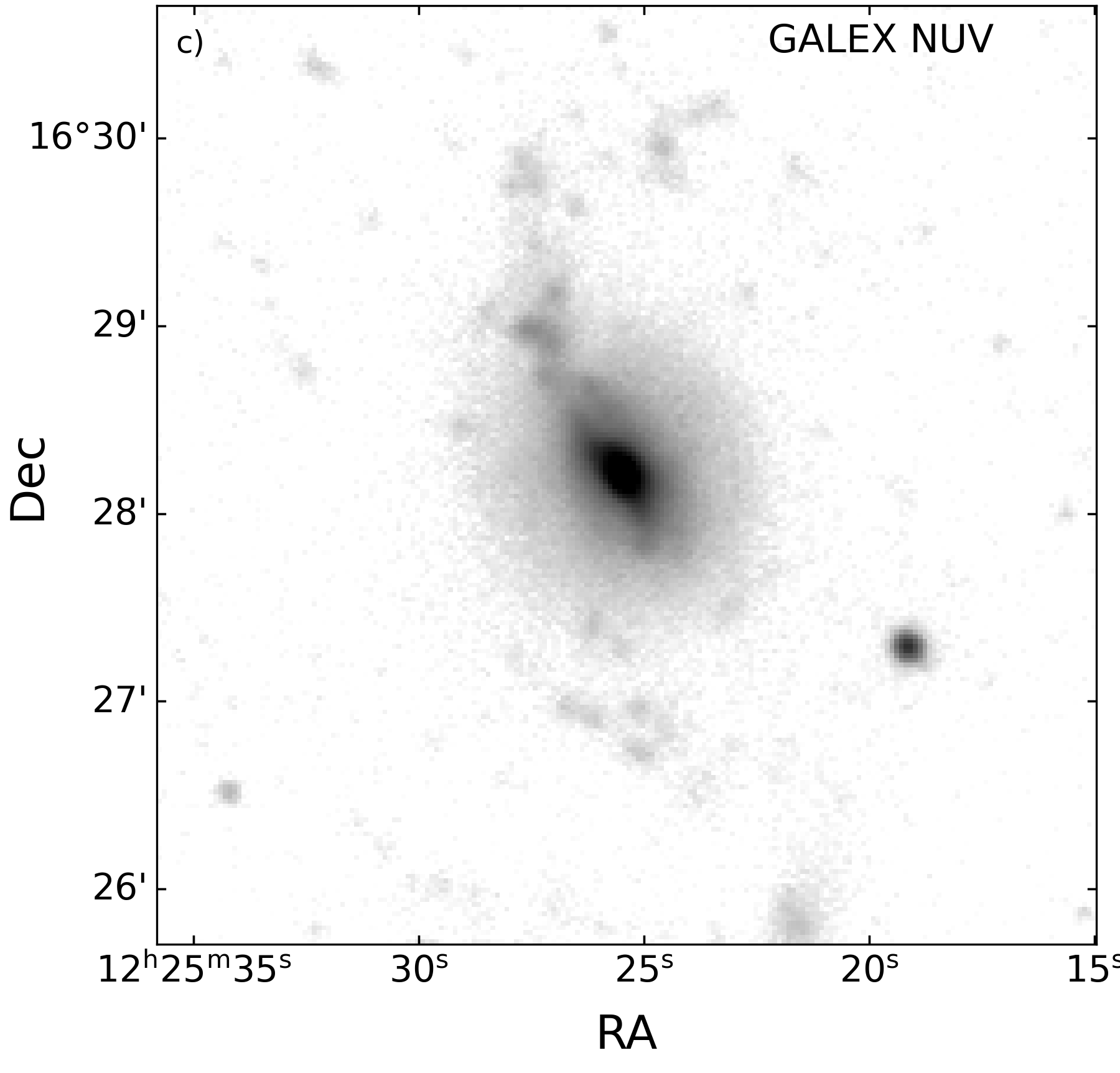}
    \includegraphics[width=0.48\textwidth]{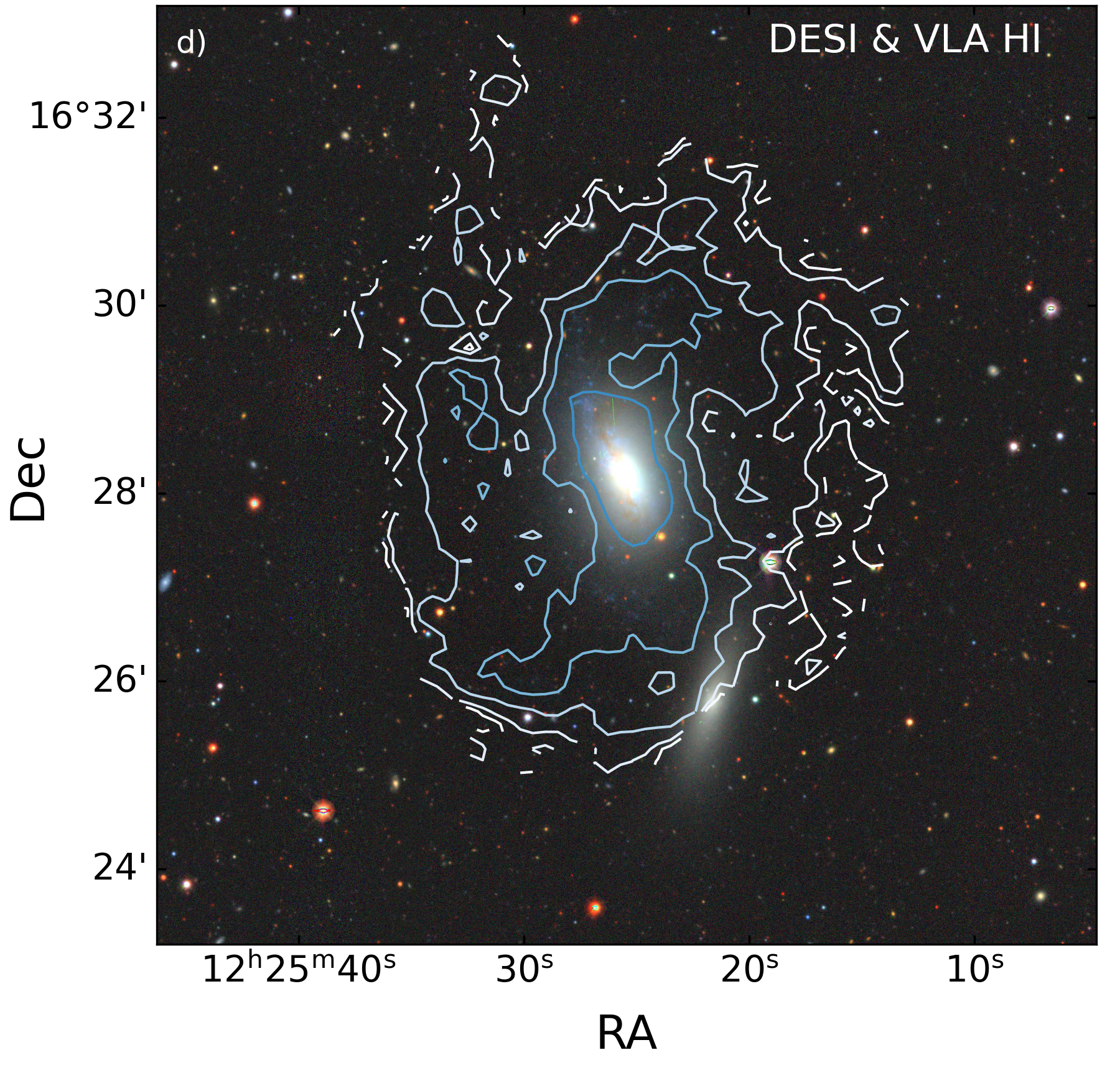}
    \caption{A historical and multiwavelength view of NGC\,4383.  a) Digitised Sky Survey $r$-band optical image (Palomar 103aE filter centred on $6400$\,\AA). North is up, east is left. b) DESI $grz$ three-colour image from the Legacy Sky Survey. The outline of our MUSE pointings is shown with thin white lines.  c) GALEX \cp{martin05} NUV image. d) \HI contours from the VLA/VIVA \cp{chung09} survey overlaid on a DESI colour image (note the larger field-of-view of this image compared to the previous ones). Levels are $\lgSigHIMsun$ = 0.1, 0.25, 0.5, 0.75, 1.
    A seemingly unremarkable object at first glance, NGC\,4383 hosts a central starburst and a prominent dust lane, a tail of \HII regions, and a massive, extended 
    \HI disc.}
    \label{fig:history}
\end{figure*}

NGC\,4383 is an intermediate-mass galaxy \cp[$\lgMstarMsun=9.44$,][]{leroy19} 
in the Virgo cluster, located $\sim1.25$ Mpc \cp[$4.3$\,deg in projection,][]{yoon17} North-East of M 87, just inside the $R_{200}$ radius of Virgo \cp[1.55 Mpc,][]{mclaughlin99, ferrarese12}.
Its position on the phase-space diagram suggests that the galaxy is on its first infall into the cluster \cp{boselli23}.
At first glance it is an unremarkable object, and in \fig{history}a, we show a Digitised Sky Survey $r$-band image, which reveals a relatively smooth and featureless galaxy. 
Morphological classifications have been uncertain, varying between an Sa galaxy with peculiar features \cp{devaucouleurs91} and an S0 \cp{trentham02}, while \ct{balzano83} first identified the central starburst.
The SFR 
from the $z=0$ Multiwavelength Galaxy Synthesis catalogue \cp[$z0$MGS,][]{leroy19} estimated from UV+IR emission is $1.02\,\Msunyr$, placing it $0.56\,$dex or $\sim1.6\sigma$
above the locus of the star-forming main-sequence defined in that paper (see their equation 19). This galaxy does not host an active galactic nucleus (see also \S~\ref{subsec:bptZ} below). 

More recent imaging surveys began to reveal the intriguing nature of NGC\,4383.
In \fig{history}b, we show a $grz$ colour image from the DESI\footnote{Dark Energy Spectroscopic Instrument.} Legacy Survey database \cp{dey19}, which reveals a clear dust lane and a very bright central component.  
To the North of the stellar body, there is an extension of blue knots that are much brighter in the near-UV (\fig{history}c) and are clear evidence of star formation occurring outside the main body of the galaxy \cp[e.g.,][]{thilker07}. 
Given NGC\,4383's infalling status into Virgo, these knots are reminiscent of star formation occasionally seen in the ram-pressure stripped tails of cluster galaxies \cp[e.g.,][]{cortese07,yamagami11,vulcani18b,poggianti19,cortese21,boselli22,george22}. 

However, the picture becomes even more complex once the neutral gas is considered. 
In \fig{history}d, we show \HI observations from the VIVA survey, which revealed that NGC\,4383 hosts one of the most extended \HI discs in Virgo, reaching over four times the optical size ($\RHI/R_\text{25$^\mathrm{{th}}$,B} = 4.2$) with a total mass of $\lgMHI = 9.46$.
The morphology of the \HI disc does not support a ram-pressure origin for the Northern star-forming knots.
Instead, the star-formation regions align with the extended higher column density \HI (the second-highest contour, $\lgSigHIMsun=0.75$, 
in \fig{history}d), suggesting that they may just be H{\sc{ii}} regions forming in the high density parts of the spiral arms of the cold gas disc. However, at this stage, a recent interaction cannot be ruled out.   

\ct{koopmann04} were perhaps the first to suggest that NGC\,4383 might host an outflow, as they noted bipolar filamentary structures in their  \HA narrow-band image, and the \HA image from the VESTIGE\footnote{A Virgo Environmental Survey Tracing Ionised Gas Emission.} \cp{boselli18} survey shows the bipolar outflow morphology clearly  \cp[see VCC 801 in figure B.1 of][]{boselli23}.
\ct{rubin99} found evidence for non-circular motions in the central regions from $\HA$ long-slit spectroscopy, while \ct{chung09} noted distinct variation in the position angle of the \HI\ disc between the gas inside the stellar component and the outer parts. 
Further, the Eastern side shows some \HI kinematic anomalies and an \HI cloud outside the disc, suggesting a possible interaction or accretion event. 
The dwarf galaxy 2.5 arcmin SW of NGC\,4383 that can be seen in \fig{history}, UGC 7504 (VCC 794), is the most likely candidate for a gravitational interaction. While \cp{koopmann04,chung09} rejected this hypothesis due to a claimed line-of-sight velocity offset of 800 \kms from NGC\,4383, more recent spectroscopic observations \cp{abazajian09,toloba14} indicate a line-of-sight velocity offset smaller than 100 \kms. Thus, at this stage, it remains unclear if, and to what degree, NGC\,4383 has experienced a recent interaction with a companion.

\section{Observations and data reduction} \label{sec:DR}
We carried out our observations with the MUSE integral field spectrograph, mounted at the Unit Telescope 4 of the Very Large Telescope at the ESO Paranal Observatory in Chile. MUSE was configured with the Wide Field Mode and nominal wavelength range set up, which ensure a field of view of $\sim 60$ arcsec$^2$, a spatial scale of 0.2 arcsec/pixel, a spectral sampling of 1.25 \AA /pixel, and a nominal spectral resolution (FWHM) of 2.5 \AA\ at 7000 \AA.
We observed NGC\,4383 with three MUSE pointings, two above and below the stellar disc targeting the extra-planar $\HA$ emission and one targeting the northern star-forming knots. 
These pointings are overlaid on a DESI optical image in \fig{history}b. 
The North West (NW) and South East (SE) extra-planar pointings were observed during 2021 in a partially completed pilot program (PI Cortese, ID: 105.208Y), while the third tail pointing was observed as part of the MAUVE large program (PIs Catinella \& Cortese, ID: 110.244E).

The total MUSE data collected constitutes 2.7h on-source time, 3x750s exposures in the NW, 6x750s exposures in the SE (2 observing blocks), and 4x750s exposures in the tail. 
The maximum seeing measured at VLT4 during the nights of our observations was a full-width-half-maximum (FWHM) of 1.3 arcsec. 
At the assumed Virgo distance the MUSE spaxels subtend 16\, pc/spaxel, while the seeing corresponds to a spatial resolution of 100\,pc. 
Data reduction was performed with \textsc{esorex} version 3.13.6 using the MUSE data reduction scripts (DRS) version 2.8.7 \cp{weilbacher20}.
We used the data reduction approach presented in \ct{emsellem22}, making use of version 2.24.7 of the \textsc{pymusepipe}\footnote{\url{https://pypi.org/project/pymusepipe/}} python wrapper for \textsc{esorex}, which includes additional procedures for exposure alignment and background flux calibration.
We describe our adopted workflow below. 

First, \textsc{pymusepipe} runs the MUSE DRS recipes up to and including \textit{muse\_scibasic} to remove the instrumental signature from the science and sky exposures and construct the sky spectrum. 
We constructed the sky continuum spectrum and identified the location of sky emission lines using the 60 per cent darkest spaxels in the dedicated sky exposures for each observing block (OB), but always re-fitted the sky line fluxes on each individual exposure (-skymethod=model).
Additionally, \textsc{pymusepipe} uses \textit{muse\_scipost} along with optical filter curves to create non-sky-subtracted flattened images. 
These images are used alongside background imaging to compute exposure alignment solutions and flux calibration adjustments. 
Background images were created using SDSS $r$-band and DESI $r$-band filters as we split our workflow's alignment and flux adjustment steps. 
In particular, we used SDSS imaging for flux calibration, but testing indicated that better alignment solutions were achieved
with the more recent and higher resolution DESI imaging. 
The SDSS background image was sourced using the SDSS DR9 mosaic tool, and the DESI image using a Legacy survey download URL\footnote{\url{https://www.legacysurvey.org/dr9/description/}}.

Alignment is the workflow's next step, and \textsc{pymusepipe} uses an optical flow algorithm to compute the pixel-by-pixel difference between flattened MUSE images and background reference imaging. 
These differences are used to compute the positional and rotational shifts required to align the MUSE exposures with the background imaging. 
We inspected the output diagnostic plots and found that the alignment solution almost always needed manual adjustment. 
There is a stable (proper motion 0.42 mas/yr) Gaia star\footnote{Gaia DR3 object 3946001070056318592} in the overlap of all three pointings, so we manually adjusted the frames to match this star in the background imaging. 
Most additional adjustments were within 4 pixels, or 0.8 arcsec.
Although we aligned on this single point source, we also noted good agreement between the stellar continuum and the locations of the Northern \HII regions between the MUSE and background imaging.

After computing the alignment solution, \textsc{pymusepipe} estimates the residual sky background and a flux calibration correction by comparing the flux in the MUSE flattened images with the reference background imaging. 
As described in \ct{emsellem22}, this method assumes that the difference between the signal-free MUSE spaxels and the sky-subtracted reference image is due to residual sky emission in the MUSE cube, while flux differences are due to flux calibration uncertainties. 

The third step runs \textit{muse\_scipost} to subtract the sky background after applying the normalisation and flux calibration derived in step 2 to the spectra in a wavelength-independent way. 
\textsc{pymusepipe} defines a global World Coordinate System (WCS) grid that encapsulates all OBs in the dataset, and datacubes are constructed from each exposure using this WCS. 
Lastly, the individual exposure cubes are mosaiced into a datacube on this common WCS grid. 
The spaxels are first sigma clipped using a 5-iteration median absolute deviation and then combined with an exposure-weighted mean.

We characterised the stellar continuum emission using the  \textsc{ppxf} \cp[penalised pixel fitting,][]{cappellari17,cappellari22} full spectral fitting routine. 
The spectra were de-redshifted, and the continuum signal-to-noise computed as $\SN_\mathrm{cont} = \frac{\mathrm{med}(f_{\lambda})}{\mathrm{med}( f_{\mathrm{err},\lambda})}$ 
in the wavelength range $5300-5500$\,\AA, where $f_{\lambda}$ is the observed spectrum and  $f_{\mathrm{err},\lambda}$ is the square-root of the MUSE variance spectrum.
Before spatially binning the spectra, we created a spatial mask to encapsulate the main stellar component of the galaxy. 
In this way, we avoided generating large spatial bins from spaxels with a weak or absent underlying continuum that would instead be dominated by sky emission.
We smoothed a map of $\SN_\mathrm{cont}$ with a 3-arcsec (15 spaxels) box filter and selected the contour corresponding to $\SN_\mathrm{cont}=1.5$ as the mask.
All spaxels within this mask were binned to $\SN_\mathrm{cont}=40$ using the Voronoi adaptive spatial binning method \cp{cappellari03}, and we logarithmically re-binned the spectra using the \textsc{ppxf\_utils.log\_rebin()} function, which resulted in a velocity scale of $57\,\kms$.  

We masked $400\,\kms$ regions around the expected positions of residual sky emission lines and all bright nebular emission lines. We tested that this approach is more reliable and robust than fitting simultaneously gas emission and stellar continuum. We also masked the $\lambda=5890,\ 5896$\,\AA\ Sodium absorption doublet feature as it can be contaminated by ISM absorption, especially in galaxies containing cool outflows. 
We used 72 MILES\footnote{Medium resolution INT Library of Empirical Spectra} \cp{vazdekis15} simple stellar population  templates with ages = (0.05, 0.08, 0.15, 0.25, 0.40, 0.60, 1.0, 1.75, 3.0, 5.0, 8.5, 13.5)\,Gyr and metallicities [M/H] = ($-$1.49, $-$0.96,$-$0.35, 0.06, 0.26, 0.4), selected to best compromise between low metallicity coverage and young populations within the MILES SAFE ranges. 
The templates were convolved with the wavelength-dependent line spread function (LSF) of the MUSE data using the parameterisation given in equation 8 of \ct{bacon17} before being logarithmically binned to the same wavelength channels as the observed spectra.

We fit the binned spectra with \textsc{ppxf} over the wavelength range $4800 - 7000$\,\AA\, including multiplicative polynomials (since we focus on emission lines only) of order 12 and four stellar kinematic moments ($V_\mathrm{stellar}$, $\sigma_\mathrm{stellar}$, h$_{3,\mathrm{stellar}}$, h$_{4,\mathrm{stellar}}$), and input the MUSE $f_{\mathrm{err},\lambda}$ spectra as the noise estimate.
We used these fits to create an emission-line-only datacube, which we used to derive emission-line data products with higher spatial resolution.  
Spaxels within Voronoi bins had their continuum subtracted by rescaling the best fit binned spectrum by the ratio of the integrated emission in the $5300-5500$\,\AA\, wavelength range, and unbinned spectra were left unaltered.

\subsection{Emission line measurements} \label{subsec:measure_emline}
Ionised gas originating from different physical components with different line-of-sight velocity distributions is often modelled with multiple Gaussian components. 
However, the coarse channel width (1.25\,\AA\ or $\sim57\,\kms$ at $\HA$) 
and spectral resolution (FWHM$\sim$ 2.54\,\AA\ or $116\,\kms$ at $\HA$ from the LSF) of MUSE means that many of these components will be blended, and while individual emission lines show evidence of non-Gaussian structure, separating them and assigning them physical meanings becomes difficult.
Instead, using the data cubes with the stellar contribution already subtracted, we constructed source masks to create line-only subcubes for the \HB, \OIIIb, \OIIIr, \HA, \NIIb, \NIIr, \SIIb, and \SIIr emission lines, which we then analysed individually. 

First, to remove any residual continuum emission, we fit a second-order polynomial baseline within a $1500\,\kms$ region around the expected rest-frame wavelength of each emission line after sigma clipping the spectrum using a two sigma threshold. 
We then created emission line masks using a method similar to \ct{leroy21a}. 
A primary mask was defined as channels with signal $\SN_\lambda>3.5$, which isolates bright emission, while the secondary mask is defined as regions of at least three channels with $\SN_\lambda>1.5$ and selects fainter emission. 
The final emission line mask was defined as all secondary mask regions, provided they also contain a primary mask region. 

The integrated flux and flux uncertainty of each emission line was measured by integrating the observed and noise spectra within the final mask region.
We also measured the line-of-sight velocity and velocity dispersion by computing the flux-weighted first and second moments of each emission line. 
Unless otherwise stated, we do not correct the velocity dispersion for LSF broadening.

\section{Physical and chemical properties of the outflow} \label{sec:outflow}
\subsection{The NGC\,4383 outflow}

\subsection{Spatial binning and spectral fitting}
\begin{figure*}
    \centering
    \includegraphics[width=\textwidth]{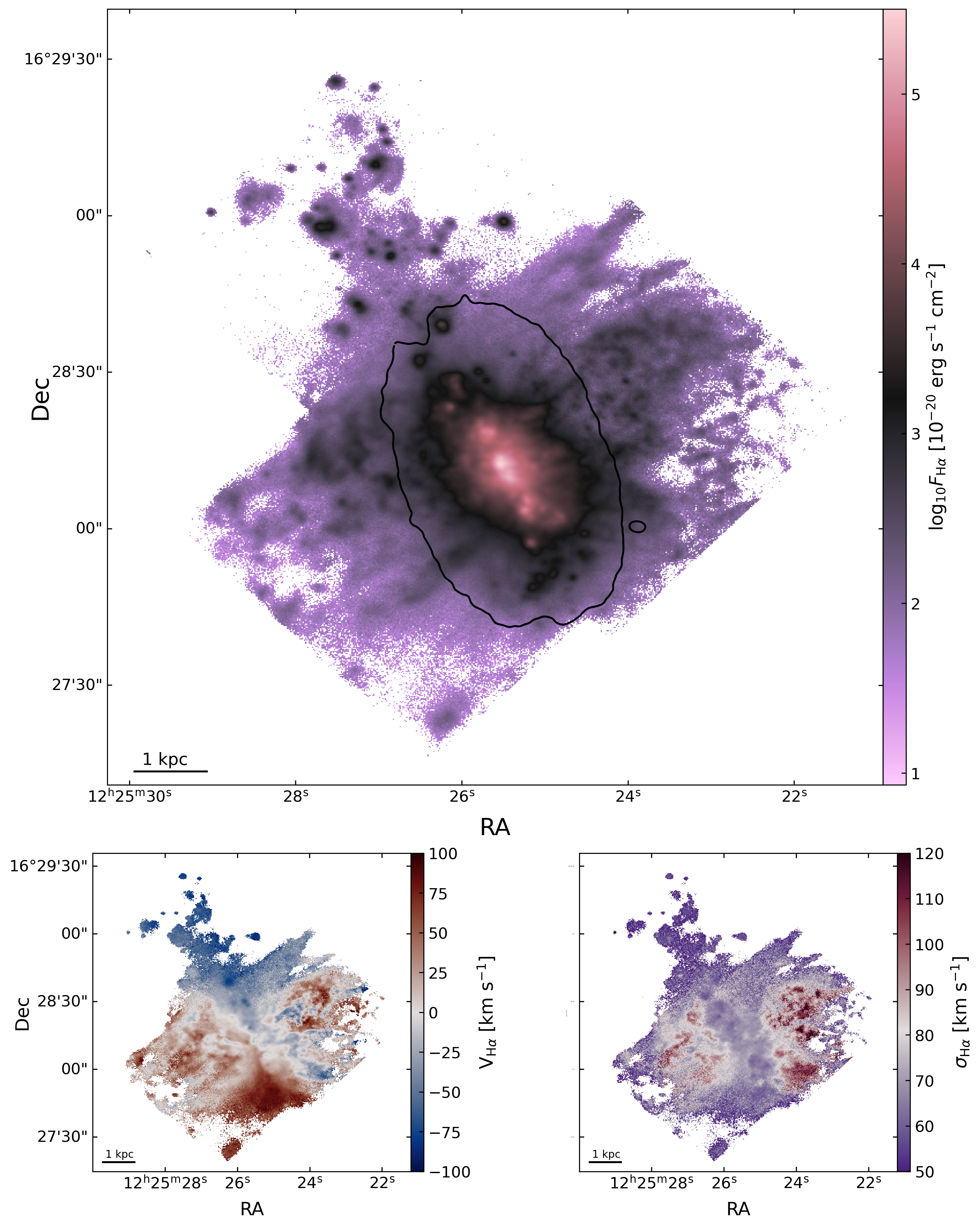}
    \caption{Derived \HA emission line properties. 
    The main panel shows our integrated \HA flux map in units of $10^{-20} \ergscm$, with the DESI $r$-band 25\ts{th}\,mag\,arcsec$^{-2}$ isophote overlaid in black. 
    The bottom two panels show the \HA flux-weighted first moment (line-of-sight velocity, left) and second moment (velocity dispersion, right).
    All panels highlight the complex and clumpy distribution and motions of the \HA gas in the galaxy. The top panel shows chimneys and shells associated with the bipolar outflow, while the bottom two panels show the multiple line-of-sight velocity components within outflow-dominated spaxels outside the galaxy's major axis.}
    \label{fig:outflow_map}
\end{figure*}

In \fig{outflow_map}, we present the integrated \HA emission line map of NGC\,4383 with no binning,  
showing that emission is detected nearly everywhere in the datacube, especially outside the optical extent of the galaxy (black contour, $r$-band 25\ts{th} mag arcsec\ts{-2}).
This image reveals the clumpy structures in the warm ionised gas, such as chimneys and shell-like 
features, seen also in previous narrow-band imaging \cp{koopmann04,boselli23}, which suggested the presence of a bipolar outflow.

\begin{figure*}
    \centering
    \includegraphics[width=\textwidth]{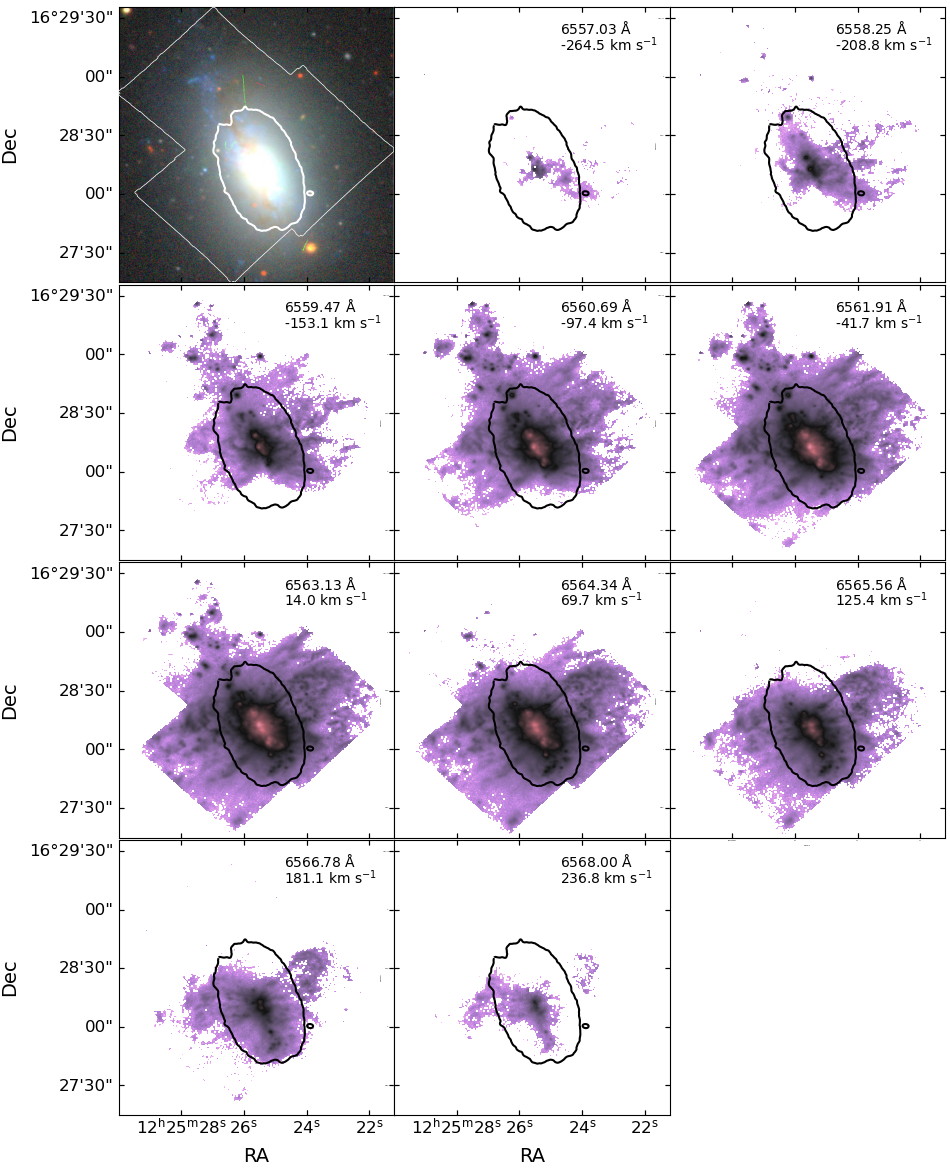}
    \caption{\HA channel maps. 
    The first panel shows a DESI three-colour image (with thin white lines indicating the outlines of the MUSE pointings), and the remaining ten show spaxel maps from the continuum-subtracted datacube, increasing in wavelength across the \HA line. 
    The white/black contour shows the  DESI $r$-band 25\ts{th}\,mag\,arcsec$^{-2}$ isophote. 
    The central wavelength and corresponding line-of-sight velocity of each spaxel map are given in the top right of each panel.}
    \label{fig:channel_map}
\end{figure*}

In \fig{channel_map}, we show channel maps of \HA emission highlighting the complex velocity structure of the outflow even with the $\sim57\,\kms$ channel width sizes. 
Blue-shifted emission appears first on the SW side of the datacube, just above the receding major axis of the galaxy. 
The outflow emission then creeps Northerly while the bright \HII regions on the approaching side of the disc begin to appear. 
Emission begins to fill the datacube as we cross the centre of the \HA line, and the chimneys and shells become visible both above and below the stellar component. 
The \HII regions fade as the receding side of the disc dominates the redshifted channels, plus the brightest shell of the outflow in the NW and the chimneys in the SE. 
These structures remain until the last of the redshifted emission traces the outflow to the East of the galaxy centre, opposite to the first channel. 

The overlap of these different velocity structures results in the complex $V_{\HA}$ map in the bottom left of \fig{outflow_map}.
The signature of the differential rotating disc is evident in the $\sim200\,\kms$ transition from approaching (blue) to receding (red) emission roughly N to S in the datacube, which traces the optical major axis. 
Off the major axis, however, the presence of a coherent rotational velocity structure breaks down. 
Instead, the gas exhibits nearly $\sim200\,\kms$ transitions within just a few spaxels, and these off-axis velocity structures agree well with the locations of chimneys and shells in the $F_{\HA}$ map. 

The complex outflow kinematics are also apparent in the  $\sigma_{\HA}$ map (bottom right). 
In the central regions along the major axis and the Northern \HII regions, the (uncorrected) velocity dispersion is close to our instrumental limit 
(LSF $\sigma\sim50\,\kms$ at $\HA$). 
Outside these regions, the velocity dispersion rises sharply and in excess of $130\,\kms$ in some spaxels, a clear indication of more complex kinematics and potentially the presence of multiple 
line-of-sight velocity components in each spaxel. 

We note that the complex velocity structure is clearer in both the $V_{\HA}$ map and the $\sigma_{\HA}$ map on NW side of the galaxy. 
Additionally, we note that computing the Balmer decrement (not shown here) reveals that extinction values are larger on the Eastern side of the galaxy. 
These observations suggest that the less-obscured NW half is tilted toward us. 
In contrast, the SE half is directed away and partly obscured by the galaxy's disc, making its velocity structure less clear and the emission more extincted. 
This geometry is also supported by the dust lanes in the main body of the galaxy, which are shifted to the SE (\fig{history}b)
We assume this orientation for the remainder of the paper.

\subsection{Outflow geometry and velocity} \label{subsec:geo}

Determining the geometry and kinematics of the outflow is useful for constraining how the outflow is being launched and the nature of its interaction with the CGM, or even the intracluster medium (ICM) in this case if the hot halo has already been swept away. 
NGC\,4383 is not a (nearly) edge-on system like other examples of well-resolved ionised gas outflows, but the galaxy's stellar component is inclined $56^\circ$ on the sky\footnote{Estimated from $r$-band isophote fitting, which can have large uncertainties. Within $0.5-3\times R_{50,r}$ the isophote ellipticity varies between $0.4-0.5$, corresponding to inclinations of $54^\circ-62^\circ$.} \cp{brown21}. 
Subsequently, each spaxel contains a mixture of disc and outflow emissions that are difficult to separate with the limited MUSE velocity resolution.
This blending of emission lines also means that we cannot reliably determine the outflow opening angle using the line-splitting approach  (except in isolated cases, which we describe below) commonly used in other works \cp[e.g.,][]{westmoquette08,leroy15,martini18}.
Instead, we determine the opening half-angle using a variation of the geometric method introduced by \ct{mcpherson23}.

\begin{figure}
    \centering
    \includegraphics[width=0.48\textwidth]{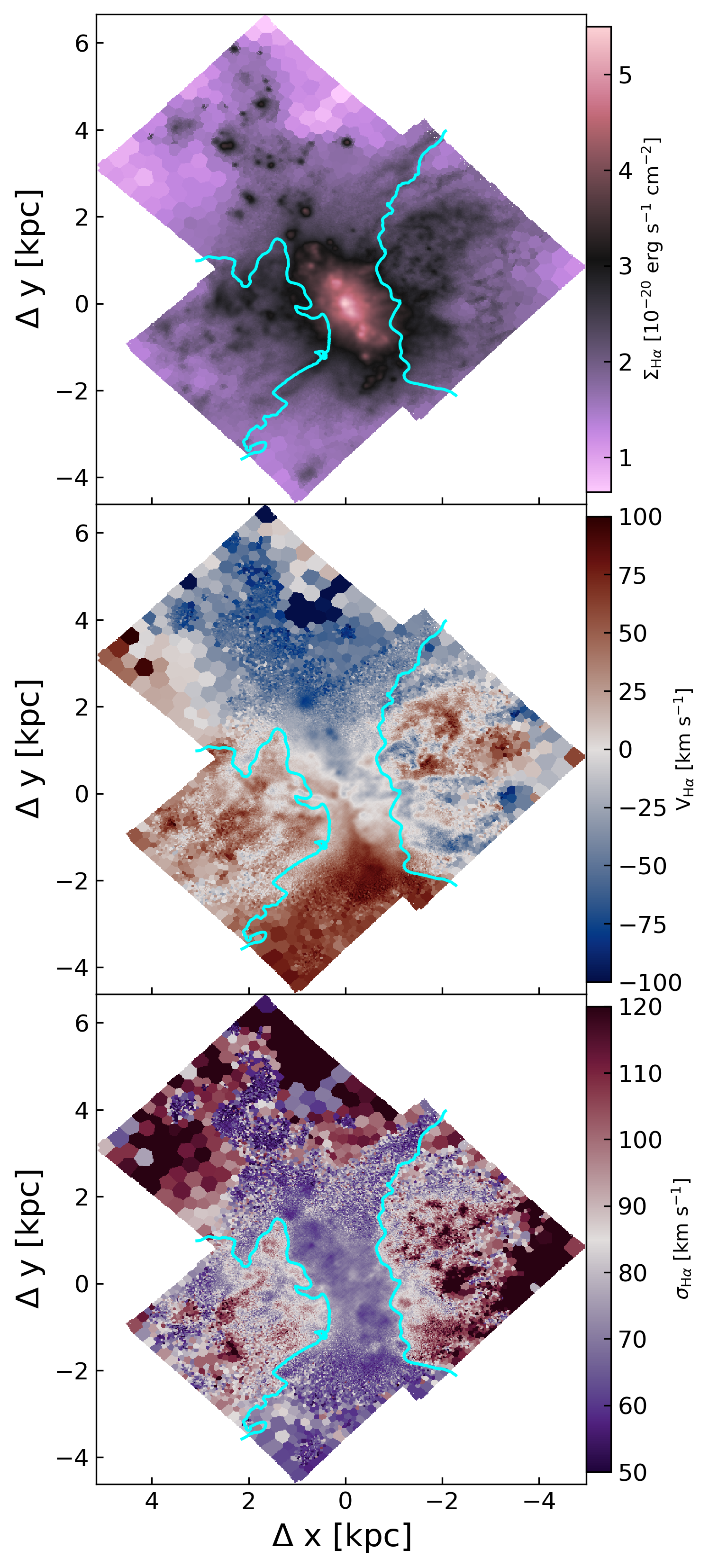}
    \caption{Definition of outflow-dominated regions. 
    Panels show the Voronoi binned \HA emission line map (top), $V_{\HA}$ (middle) and $\sigma_{\HA}$ (bottom). 
    The cyan contour in all panels corresponds to  $\sigma_{\HA}=77.5$\,\kms, which we use to separate the regions dominated by outflow emission.
    The cyan contour encloses the shells and clumpy structures associated with the outflow in the $F_{\HA}$ map, and the associated regions of anomalous $V_{\HA}$  and elevated $\sigma_{\HA}$.}
    \label{fig:outflow_defn}
\end{figure}

First, we binned the datacube to increase the $\SN_{\HA}$ above the native-resolution data products. 
We applied the Voronoi binning method with a target of $\SN_{\HA}=10$, assuming non-detected spaxels had $F_{\HA} = 0.5 F_{\mathrm{err}, \HA}$ (i.e., a maximum of 20 spaxels per bin).
Spectra from the continuum-subtracted cube were combined using these new Voronoi bins, and the properties of the emission lines were re-measured in each bin using the same procedure described above in \S\ref{subsec:measure_emline}.
We used elevated $\sigma_{\HA}$ regions to identify spaxels dominated by outflow emission. 
We overlaid contours of $\sigma_{\HA}$ smoothed with a 3-arcsec square kernel (15 pixels) on the binned $\sigma_{\HA}$ map. 
Contours enclosing the outflow regions were visually selected in NW and SE regions using the conditions that 1) they reasonably enclosed the off-plane regions of elevated velocity dispersion and 2) they did not cross through the disc. 

This process is demonstrated in \fig{outflow_defn}, with outflow-enclosing contours shown in cyan, corresponding to a velocity dispersion of $\sigma_{\HA} = 77.5\,\kms$.
We note that due to the increased size of the spatial bins, some of the $\sigma_{\HA}$ values can be artificially elevated. 
Regardless, our choice of contours is supported by their enclosing the clumpy, off-plane flux distribution and the regions of anomalous $V_{\HA}$. Our metallicity analysis in \S\ref{subsec:bptZ} also supports our choice. 

The outflow-dominated emission extends to the corners of the datacube, allowing us to place a lower limit on the extent of the outflow above the plane of the disc. 
The most distant spaxel in the NW outflow is 4.91 projected kpc (61.4 arcsec) above the galaxy's major axis. 
Correcting by $1/\sin(i)$ assuming the outflow is perpendicular to the disc, the detected outflow emission extends at least 5.94\,kpc above the galaxy. 

\begin{figure*}
    \centering
    \includegraphics[width=\textwidth]{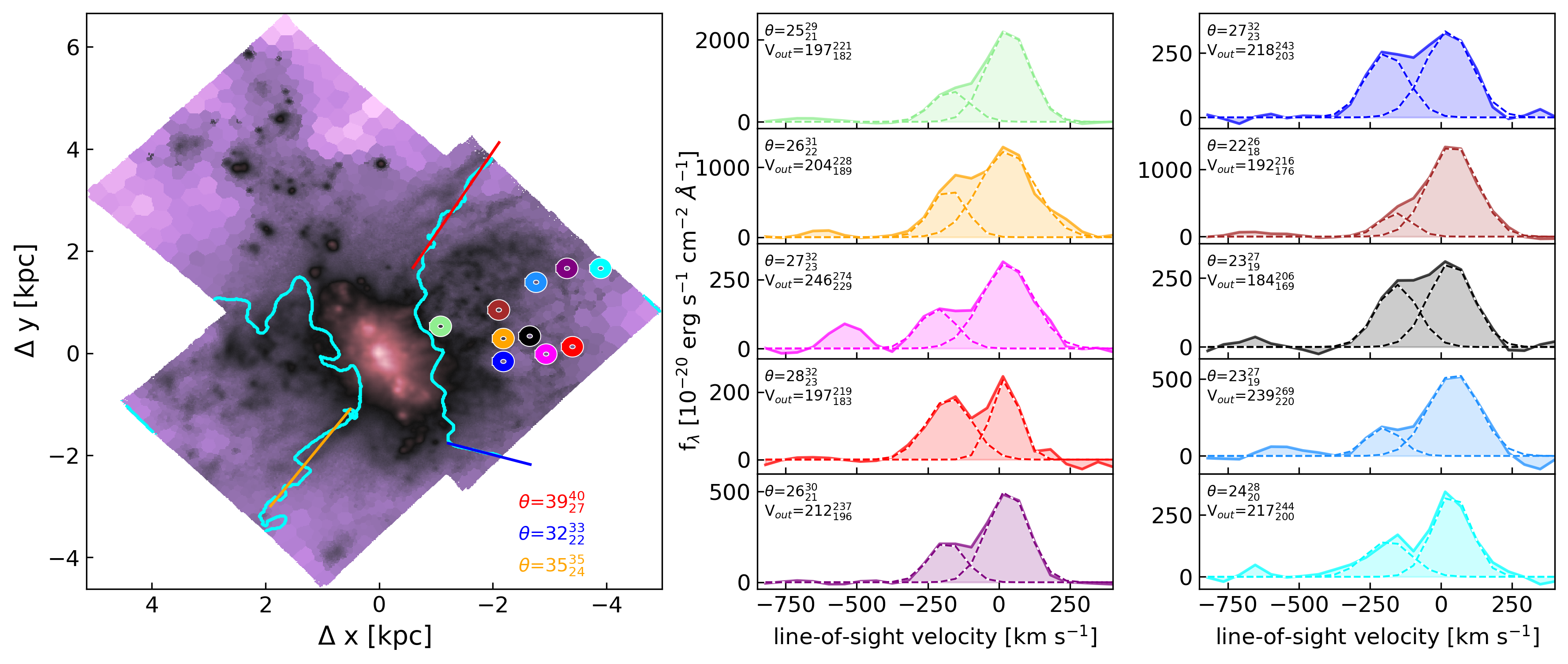}
    \caption{Outflow opening half-angle ($\theta$) measurements. Left: Contours defining the outflow-dominated emission overlaid on the binned \HA flux map. Red, blue, and orange lines show the linear fits used to estimate $\theta$. The inclination-corrected estimates for $\theta$ are given in the bottom right corner in their corresponding colour, all of which agree well.  
    Right: Spectra that exhibit line splitting, extracted from the like-coloured apertures (enlarged by $3\times$) in the NW region of the outflow in the left panel. Dashed lines represent the two Gaussian fits to each spectrum, and the derived outflow velocity and opening half-angle are quoted in the top left corner of each panel.
    In both panels, variations in $\theta$ and $v_\mathrm{out}$ due to a change in inclination of $\pm5^\circ$ are given as superscripts and subscripts
    Across all methods, $\theta$ is constrained to $22^\circ-39^\circ$, with outflow velocities in the range $184-246\,\kms$.}
    \label{fig:OA}
\end{figure*}

To measure the outflow opening half-angle, we fit straight lines $(y=mx+c)$ to the edges of the NW and SE outflows after rotating the galaxy by its stellar PA=$107.5^\circ$, which we show in the left panel of \fig{OA}. 
We do not fit the NE contour due to being poorly constrained at the cube's edge. 
Assuming the outflow is launched perpendicular to the plane of the stellar disc, the opening half-angle of the outflow is then given by $\theta = 90-\arctan(m/\sin(i))$. 
We find a range of $\theta=32-39^\circ$ and report the estimate for each fit in the bottom right corner of \fig{OA}, with the super- and sub-scripts corresponding to the uncertainty originating from a $\pm5^\circ$ variation around the assumed inclination of $i=56^\circ$, and note the good agreement between all estimates.

As an independent check on this estimate, we searched the continuum-subtracted cube at native resolution for individual cases of line splitting.
We found some isolated cases of line splitting, typically in regions where the average line-of-sight kinematics (\fig{outflow_map}) transitions between approaching and receding, and selected examples close to the minor axis to remove influence of any residual rotational velocity carried by the ejected gas \cp[visible parallel to the minor axis, e.g.,][]{leroy15,mcpherson23} on the centroids of lines.
These cases are shown in the right panels of \fig{OA}, with the colours showing their spatial extraction locations in the left panel. 
To determine more accurate centroids, we fit two Gaussian components to the spectra, and using the simple geometric arguments presented in Appendix \ref{app:app1}, the outflow opening half-angle is estimated as $\theta=22-28^\circ$.
On average, these values are smaller than calculated from straight line fits presented earlier but consistent within the reported uncertainties, giving us confidence that in these cases of line-splitting, there must be a minimal contribution from disc emission and the spatial mask defined above is reliably separating disc and outflow-dominated regions.
Taking the average value from the two methods, the opening half-angle of the outflow is $\theta=25-35^\circ$.

Using the kinematics of the line-splitting, we estimated an intrinsic (i.e., corrected for inclination) outflow velocity in the range $v_\mathrm{out}=184-246\,\kms$ (average $210\,\kms$), with no clear trend with increasing height above the disc. 
These values apply to the bulk of the gas in a given line of sight, so we also estimate the maximum outflow velocity in a given sight line as $v_\mathrm{max} = v_\mathrm{out} + 2\sigma$ where $\sigma$ is the LSF broadening-corrected standard deviation of the fit to the blue-shifted component. 
Using this approach, we find maximum outflow velocities in the range $v_\mathrm{max}=250-340\,\kms$ (average of $300\,\kms$).

These values can also provide some insights into whether the outflowing gas is primarily energy- or momentum-driven. Using our extinction-corrected $F_{\HA}$ emission line map, we computed the SFR surface densities within $R_{50,r}$, and found a range of values of $\log\Sigma_\mathrm{SFR} [\Msunyr\mathrm{kpc}^{-2}] = [-1.5,0.9]$ with an integrated value of $\lgSigSFR=-0.27$. Simply assuming that our maximum ($340\,\kms$) and minimum ($250\,\kms$) $v_\mathrm{max}$ are driven by the largest and smallest $\SigSFR$ respectively, the NGC\,4383 outflow agrees with the average trend of the data points in Figure 4 of \ct{reichardtchu22a}, making it more consistent with an energy-driven outflow accelerated by supernovae feedback. For example, for $\log\Sigma_\mathrm{SFR} [\Msunyr\mathrm{kpc}^{-2}]\sim0.9$, maximum velocities expected the radiation pressure-driven outflows exceed our observed values by a factor of a few. However, as the average $\Sigma_\mathrm{SFR}$ and outflow velocities measured for NGC4383 lie in the region of parameter space where the momentum- (\citealp{murray11}) and energy-driven (\citealp{chen10}) models overlap, at this stage we cannot exclude that radiation pressure is playing a role here.

To test if any of the outflowing gas will escape the galaxy, we adopted the simple, spherically symmetric isothermal sphere approximation to estimate the escape velocity, $v_{\mathrm{esc}}\approx3v_\mathrm{circ}$ \cp[e.g.,][]{veilleux20}. 
The integrated \HI spectrum of NGC\,4383 is regular and shows no signs of disturbance, so we use the circular velocity $v_\mathrm{circ}=140\,\kms$ calculated using half spectral profile width measured at 20 per cent of the peak flux, and corrected for inclination.
Considering our $v_\mathrm{out}$ estimates, the bulk of the gas in the outflow likely will not escape the galaxy's potential. 

Last, we note that solving for $y=0$ from our linear fits to the NW outflow, we derive an outflow base width of 1.70\,kpc (21.2 arcsec), which suggests a wide-based outflow better described with a frustum geometry than a cone.
Interestingly, the three brightest central \HII regions span a major axis length of 1.96\,kpc (24.5 arcsec), close to the derived outflow base. 
Both these values are close to twice the $r$-band effective radius, $2\times R_{50,r} = 1.77\,$kpc (22.2 arcsec).
Comparatively, the central starburst in M\,82 has diameter $\sim1\,$kpc \cp{kennicutt98} while \HI kinematic modelling of M\,82 suggests an outflow base diameter of $\sim2\,$kpc \cp{martini18}.

\subsection{Ionisation and metallicity of the disc and outflow} \label{subsec:bptZ}
\begin{figure*}
    \centering
    \includegraphics[width=\textwidth]{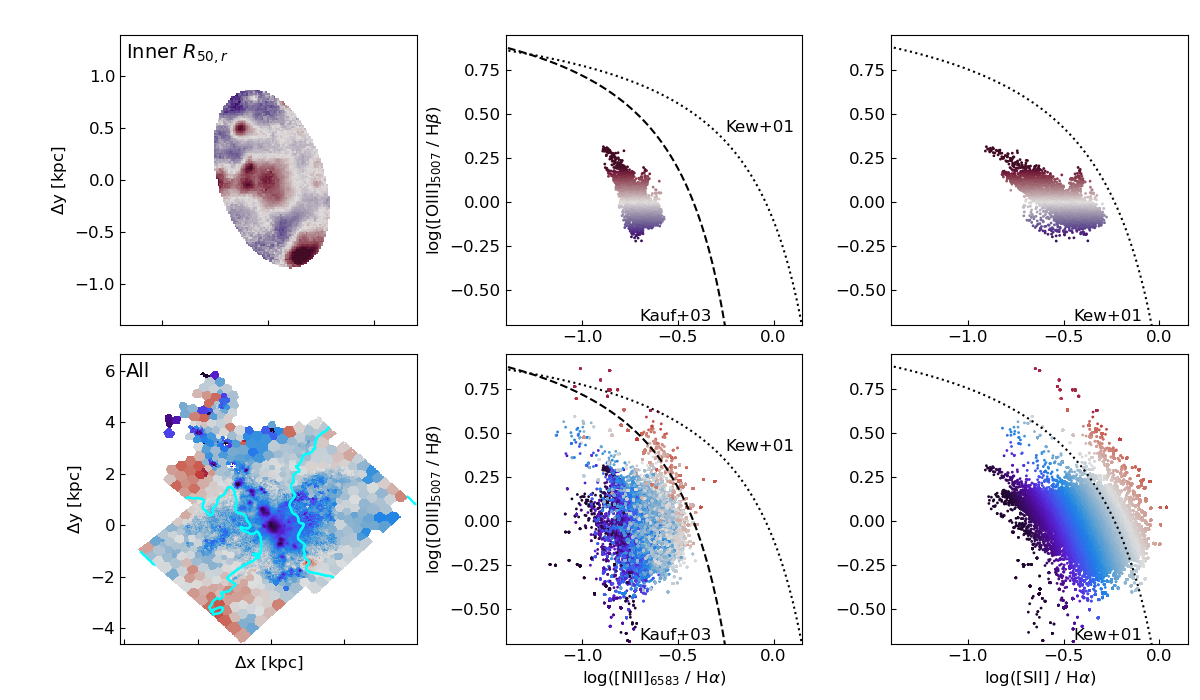}
    \caption{Spatially resolved BPT diagrams. The left column shows the spatial distribution of spaxels in the galaxy centre (within $1\,R_{50,r}$, top row), and the entire galaxy (bottom row). 
    The middle and right columns show \NIIr/\HA and \SII/\HA vs \OIIIr/\HB BPT diagrams for the spaxels in each selection, respectively.
    In the top row, spaxels are coloured by their \OIIIr/\HB ratio, while the bottom row is coloured by their \SII/\HA distance from the \ct{kewley01} demarcation line.
    The galaxy's centre shows a clear contrast between high-ionisation \HII regions and the more diffuse emission. 
    The core of the outflow in NGC\,4383 is ionised by star formation, while the contribution from shocked gas becomes more important outside the core as the spaxel colours transition from blue through to white.}
    \label{fig:BPTs}
\end{figure*}

Using the outflow masks defined in the previous section, we can compare the physical conditions in the ISM between the disk- and outflow-dominated regions.
To ensure sufficient signal-to-noise in all relevant lines, we Voronoi-binned the datacube again, this time to a minimum $\SN_{\HB}=10$, and re-measured the emission line values in each bin. 
Additionally, we corrected the emission lines for extinction using a \ct{cardelli89} extinction curve with $R_V=3.1$ and assuming an intrinsic Balmer line ratio of $F_{\HA}/F_{\HB}=2.83$.
In \fig{BPTs}, we show \OIIIr/\HB versus \NIIr/\HA and \SII/\HA BPT \cp{baldwin81} diagrams, highlighting the bright star-forming regions in the centre (within $1\,R_{50,r}$) in the top row, and the entire galaxy in the bottom row.
Spaxels in the top row are colour-coded by their $\OIIIr/\HB$ ratio, while the bottom row they are coloured by their \SII/\HA distance from the \ct{kewley01} demarcation line.

The BPT diagrams of the central region show structure, suggesting that the most SW \HII region has a larger ionisation parameter \cp{kewley19} than the core and  Northern regions.
Spaxels across the galaxy also show an interesting \SII/\HA structure. 
The core of the outflow in the NW and SE regions is ionised by star formation as their \SII/\HA and \OIIIr/\HB ratios are bluer and sit to the left of, and below, the \ct{kewley01} demarcation line. 
Further from these star formation-dominated regions of the outflow the \SII/\HA increases, but still generally consistent with photo-ionisation with only the edges of the outflow hinting towards a possible ionisation via shocks.

\begin{figure*}
    \centering
    \includegraphics[width=\textwidth]{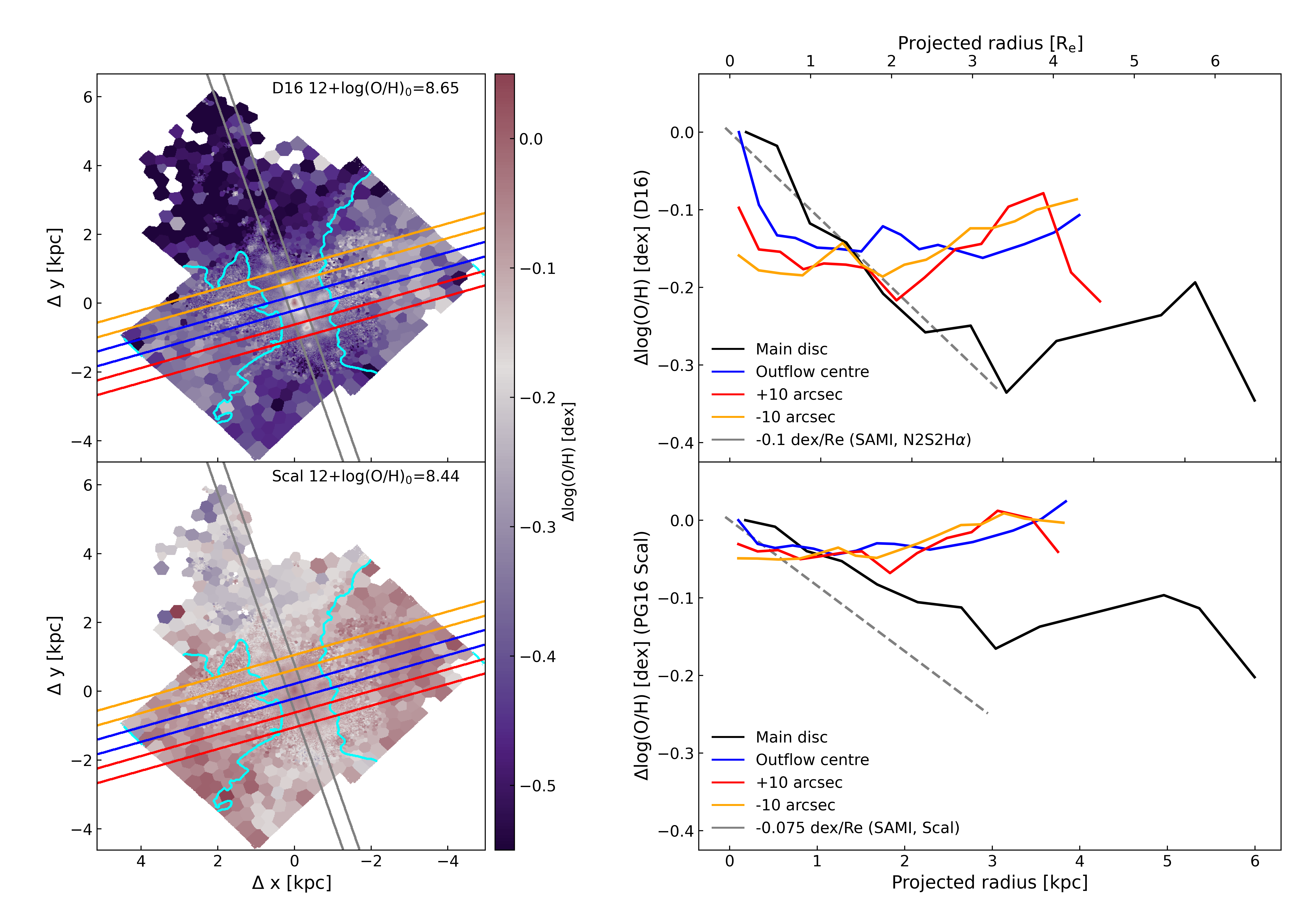}
    \caption{Metallicity gradients in the disc and outflow. The top two panels use the \ct{dopita16} metallicity calibration, while the bottom two use the \ct{pilyugin16} `Scal' calibration. Left: spatially resolved \TWlogOH map of NGC\,4383, normalised to the central metallicity, quoted in each panel's top right. Cyan contours highlight the outflow-dominated regions. Right: radial metallicity gradients aligned with the major axis (black line, grey slit in left panel), the minor axis (blue) tracing the outflow centre, the outflow centre and $\pm10$\,arcsec (red and orange lines and slits). All radial profiles are normalised to the central metallicity, and we show the typical radial gradient for a $\lgMstarMsun \approx9.5$ galaxy from \ct{poetrodjojo21} for each calibration with a grey line.
    The main result of this figure is that the metallicity in the outflow remains elevated compared to the disc but is not more metal rich than the galaxy's centre.}
    \label{fig:dlogOH}
\end{figure*}

We use emission line ratios of the star formation-dominated regions to estimate the gas-phase oxygen abundance of the ISM and quantify the enrichment due to the outflow. 
We adopt two different metallicity calibrations to ensure the robustness of our results. 
The first is the calibration of \ct{dopita16}, which uses the \NIIr, \HA, and \SII lines. 
The second is the `Scal' calibration from \ct{pilyugin16}, which uses the \OIII, \HB, \NII, and \SII lines, and we use $\OIII = 1.33\OIIIr$ and $\NII = 1.34\NIIr$. 
To highlight the spatial variation in $\logOH$, we compute the difference from the maximum value inside a 2-arcsec square at the galaxy's centre for each calibration, which is $\TWlogOH=8.65$ and 8.44, respectively. 

In the left panels  \fig{dlogOH}, we show a spatial map of $\Delta\logOH$ with the outflow contours overlaid, for all spaxels regardless of their ionisation. 
Clearly, the metallicity of the gas decreases with increasing distance from the central starburst. 
This decrease, however, is smaller in the direction of the outflow-dominated regions compared to the plane of the disc (aside from a few bright \HII regions), highlighting the metal-enriched nature of the outflowing gas, and this trend is apparent in both metallicity calibrations. 
Note that our contours defining the outflow-dominated regions derived from the $\HA$ kinematics agree well with the regions of elevated $\logOH$, further supporting the robustness of our spatial definition of the outflow. 

To quantify these differences, we computed radial profiles using  5-arcsec wide pseudo-slits in different directions on the sky. 
The disc-plane metallicity is traced with a slit along the major axis, and to quantify the outflow regions, we use slits parallel to the minor axis with major axis offsets of  0 and $\pm10$ arcsec.
Computing these profiles we only use the star-formation dominated spaxels, defined as those to the left of the \ct{kewley01} demarcation line in the \SII/\HA BPT diagram.
In the right panels of \fig{dlogOH}, we show profiles of $\Delta\logOH$ in 1-arcsec bins of circular, sky-plane radii expressed in kpc and $R_{50,r}$. 
Confirming what is visually apparent in the left panels of \fig{dlogOH}, $\Delta\logOH$ remains larger in the outflow-dominated regions while it drops sharply with radius within the disc. 
In both panels, we show the average metallicity gradient for a galaxy with a similar stellar mass to NGC\,4383 from \ct{poetrodjojo21} with a grey line, and the disc-slit radial profile agrees with these average gradients for both prescriptions. 
At 3 $R_{50,r}$ the difference in metallicity is $\sim0.15$ dex between the disc- and outflow-dominated regions for the \ct{dopita16} calibration, while the \ct{pilyugin16} `Scal' calibration has a smaller difference of $\sim0.09$ dex. 
This difference is roughly consistent with the factor of 0.53 in relative variation between the two calibrations \cp{poetrodjojo21}. 
Interestingly, in no location do we see that the metallicity of the outflow region is larger than the centre of the disc. 
Beyond 3 $R_{50,r}$, the metallicity in the disc shows evidence of flattening with radius, which could be a signature of gas inflow \cp{bresolin12} that could be feeding the central starburst.

Last, we used the electron density-dependent $\SII$ doublet to constrain the effect of the outflow on the density of ISM. 
We adopt the parameterisation from \ct{kewley19} assuming a metallicity of $\TWlogOH=8.53$, which is roughly consistent with the value in the galaxy's centre.
The electron density $n_\mathrm{e}$ is only measurable in the central, brightest region of the galaxy, and the maximum line ratio of $\SIIb/\SIIr=1.67$ implies a maximum density of $n_\mathrm{e}\sim382\, \mathrm{cm}^{-3}$. 
However, within 280\,pc (2.5\,arcsec), the line ratio drops below the low-density limit of $42\, \mathrm{cm}^{-3}$ ($\SIIb/\SIIr<1.4$) and remains below this value throughout the rest of the datacube.

These results paint the picture of an outflow that is strongly star-formation driven, is only moderately enriched compared to the disc, and causes a rapid decrease in ISM density outside the central, outflow-launching starburst.

\subsection{Mass outflow rate and mass loading factor}
Mass outflow rates are difficult quantities to estimate, particularly when using ionised gas tracers, as we need to make several assumptions about the gas temperature and density and the spatial distribution of the gas within the assumed emitting volume. 
Further, the ionised gas phase does not carry the majority of the mass in the outflow, which is instead carried by the cooler phases \cp[\HI and molecular gas; e.g.,][]{martin-fernandez16,kim20,kim20a,rey23}.
However, the current \HI and molecular gas data are too low spatial resolution to add additional constraints to the outflowing gas mass. 
Thus, while we calculate ionised gas masses and outflow rates for comparison with previous works, we preface this section with the caveat that the derived numbers can have order-of-magnitude uncertainties. 

The mass of ionised gas within the outflow-dominated regions can be calculated as 
\begin{equation}\label{eq:mass}
    M = \mu\,m_\mathrm{H}\,n_\mathrm{p}\,\delta\,V,
\end{equation}
where $m_\mathrm{H}$ is the Hydrogen mass and $\mu=1.4$ accounts for the Helium abundance, $n_\mathrm{p}$ is the number density of ionized hydrogen atoms, $\delta$ is the assumed volume filling factor of the gas, and $V$ is the volume being considered. In the following, we assume $n_{p}$=$n_{e}$, meaning that the free parameters are $n_{e}$, $\delta$ and $V$.

As our spatial coverage of the outflow is not uniform, we make the simplifying assumption that the volume $V$ is given by the area of a pixel or Voronoi bin multiplied by a line-of-sight emission path, $V=A\times l$.
Although the path length through a cone varies as a function of its location and inclination, as we are averaging all the outflow emissions, we assume an effective $l$ equal to the diameter of a cylinder with the same volume as a biconical frustum. 
The base of the outflow estimated in \S\ref{subsec:geo} has a radius of 0.85\,kpc, and the maximum height of the frustum above the disc traced by the NW outflow region is 5.94\,kpc. 
With this geometry (see Appendix \ref{app:app1}), the radius of the frustum at this height is 4.28\,kpc. 
The diameter of the cylinder with a volume equal to this frustum is $5.5$\,kpc. 
and the total area of our outflow-dominated regions is $24.4\,\mathrm{kpc}^2$, giving an effective volume of $V=134\,\mathrm{kpc}^3$.

Estimating the filling factor is more challenging. Milky Way observations suggest that the diffuse ISM has a volume filling factor of $~\sim0.2$ \cp{reynolds91}, while \HII regions in local galaxies have $0.01-0.1$ \cp{kennicutt84}. Studies of the M~82 outflow report filling factors in the range $10^{-3}-0.1$ \cp{yuan23} and $10^{-4}-10^{-3}$ \cp{xu23}.
Simulations by \ct{kado-fong20b} and \ct{rathjen23} find that the warm ionised gas consistently has a volume filling factor of $0.04-0.1$.
As the \HA emission in outflows arises from the mixing and cooling of the hot, volume-filling phase with entrained, cooler ISM, we expect the observed \HA emission to arise from more diffuse gas on the boundaries of cold clouds and in their mixed tails \cp[e.g.,][]{sparre19,gronke22, fielding22}. 
Thus, here we assume a somewhat intermediate volume filling factor of $\delta=0.01$, i.e., \HA emitting gas makes up 1 per cent of the considered volume. However, we acknowledge that this is a large uncertainty in our mass estimate. 

With assumptions on the outflow volume and gas filling factor set, one of the most common approaches used to measure mass outflow is to take advantage of the fact that the H$\alpha$ emission measure of the outflowing gas is linked to both the average electron density along the line of sight and the H$\alpha$ surface brightness of the gas outflow:
\begin{equation}\label{eq:EM}
    \mathrm{EM}\,[\mathrm{cm}^{-6}\,\mathrm{pc}] = \int n_\mathrm{e}^{2}\ \mathrm{d} l = <n_{e}^{2}> L = 4\pi \frac{\Sigma_{H\alpha}}{\alpha_{23} h\nu_{H\alpha}}
\end{equation}
where $<n_{e}^{2}>$ is the average square electron density along the line-of-sight emission path ($L$) of the outflow, $\Sigma_{H\alpha}$ is the typical H$\alpha$ surface brightness of the outflow and $\alpha_{23}$ is the effective re-combination coefficient, assuming an electron temperature of 10,000 K and case B recombination (1.17$\times$10$^{-13}$ cm$^{3}$ s$^{-1}$, \citealp{osterbrock89}). We can thus use Eq.~\ref{eq:EM} to estimate the average square electron density of the outflow:
\begin{equation}\label{eq:ne2}
<n_{e}^{2}> [\mathrm{cm}^{-6}] = 4\pi \frac{\Sigma_{H\alpha}}{\alpha_{23} h\nu_{H\alpha} L} = 4.87 \times 10^{17} \frac{\Sigma_{H\alpha}}{L}
\end{equation}
where $L$ is in parsec. With an average $\Sigma_{H\alpha}$ of the outflow is 1.37$\times$10$^{-16}$ erg cm$^{-2}$ s$^{-1}$ arcsec$^{-2}$ and $L$=5.5 kpc, we find $<n_{e}^{2}>$=1.2$\times$10$^{-2}$ cm$^{-6}$. It is interesting to note that $<n_{e}^{2}>$ is linked to $n_{e}$ of the medium (assumed to be constant) via the filling factor: $\delta$=$<n_{e}^{2}>$/$n_{e}^{2}$=$<n_{e}>$/$n_{e}$. As such, the value of $\delta$ here assumed would imply an electron volume density $\sim$1.1 cm$^{-3}$. This value is consistent with the extrapolation to low SFR surface densities of the relation obtained by \cite{xu23b} linking $\rm \Sigma_{SFR}$ to n$_{e}$ (i.e., $\sim$2 cm$^{-3}$).

We now have all the elements to calculate the outflow mass, and we can re-write Eq.~\ref{eq:mass} in terms of $<n_{e}^{2}>$, where the main free parameter is the filling factor:
\begin{equation}\label{eq:mass2}
    M = \mu\,m_\mathrm{H}\,V\,\sqrt{<n_{e}^{2}>\delta} = 5.1\times10^{7} \Big(\frac{V}{134\, \mathrm{kpc}^3}\Big) \Big(\frac{\delta}{0.01}\Big)^{0.5} ~[M_{\odot}]
\end{equation}
We note that, when the gas electron density is known (e.g., using the $\SIIb/\SIIr$ line ratio technique),  Eq.~\ref{eq:mass2} is usually re-written in terms of just the H$\alpha$ surface brightness and the electron density, as the filling factor can be fully derived from the observations (e.g., \citealp{xu23}). In our case, the upper limit in electron density $<42\, \mathrm{cm}^{-3}$ makes this approach not stringent enough to give useful constraints on the filling factor ($\delta>$7$\times$10$^{-6}$) and the outflow mass  ($M$>0.13$\times$10$^{7}$ M$_{\odot}$).

Assuming a constant outflow velocity and using the maximum observed height above the disc, the integrated mass outflow rate is 
\begin{equation}
    \dot{M}_\mathrm{out}[\Msunyr] = 1.76\, \Big(\frac{M}{5.1\times10^7\,\Msun}\Big)\, \Big(\frac{h}{5.94\,\mathrm{kpc}}\Big) \,\Big( \frac{v_\mathrm{out}}{200\, \kms}\Big)
\end{equation}
Once combined with the integrated UV+IR $\mathrm{SFR}=1.02\,\Msunyr$ of NGC\,4383, the mass loading factor is $\eta_\mathrm{M} = 1.7$. If we use instead the SFR within the effective radius ($R_{50,r}$) measured from our extinction-corrected \HA map, $\mathrm{SFR}=0.76,\Msunyr$, we obtain $\eta_\mathrm{M} = 2.3$. 
Considering the uncertainties involved, these values are comparable to recent estimates of the ionised gas mass loading in nearby galaxies, and consistent with literature scaling relations. For instance, for M\,82 \ct{yuan23} find
$\eta_\mathrm{M, M82}=1.2$ while \ct{xu23} reports $\eta_\mathrm{M, M82}\sim0.5$, \ct{heckman15} find an average $\eta_\mathrm{M}=2$ in their sample of local galaxies, and the SFR-based scaling relation of \ct{xu22a} predicts a value of $\eta_\mathrm{M}=4.2$ for NGC\,4383, although they are more sensitive to neutral gas.

\section{Discussion and conclusions} \label{ref:concl}
In this paper, we have presented the first results from the MUSE and ALMA Unveiling the Virgo Environment (MAUVE) survey, a new VLT/MUSE large program targeting Virgo cluster galaxies to understand the impact of environmental mechanisms on galaxy evolution. 
We have not used the ALMA data from the VERTICO survey \cp{brown21} in this work, instead, we demonstrated the excellent MUSE data quality and focused our analysis on NGC\,4383, a gas-rich peculiar galaxy that hosts a massive  outflow. 
Our main results are as follows:
\begin{itemize}
    \item NGC\,4383 hosts a starburst-driven outflow. The ionised gas outflow has an opening half-angle of $\theta=25-35^\circ$ and extends at least to the spatial limit of our data, i.e. at least 5.9\,kpc above the disc. 
    \item The outflow emission has an average velocity of $v_\mathrm{out}\sim210\,\kms$, with a maximum outflow velocity of $v_\mathrm{max}\sim300\,\kms$, which agrees with expectations from empirical relations based on the galaxy's integrated stellar mass and SFR (see below). 
    \item The outflow shows an interesting ionisation structure, where the base and core of the cone are clearly consistent with photoionisation, but the line ratios gradually get closer to a shock-ionised origin towards the edge of the cone. 
    \item Using strong line metallicity calibrations, we find that the outflow is metal-enriched compared to the disc but has lower metallicity than the central starburst region. 
    \item The mass outflow rate of the ionised gas integrated over the outflow-dominated regions is estimated to be $1.9\,\Msunyr$, corresponding to a mass-loading factor of $\eta_\mathrm{M}\sim1.7-2.3$.
    \end{itemize}
While NGC\,4383 shares some similarities with the archetypal outflow M~82, it differs in other ways, making it a useful addition to the list of known outflows. 
Perhaps the property that sets it most apart from other systems is its massive, regular \HI disc that extends over four times its $B$-band optical extent \cp{chung09}. 
This \HI disc does not show signs of strong interaction,
unlike the known interacting systems M\,82 and NGC\,839, making it hard to discern what triggered the central burst of star formation that is driving the outflow.
\ct[][ see Fig. 21]{chung09} note the presence of kinematic anomalies on the Eastern side of the \HI disc and suggest that perhaps NGC\,4383 has recently undergone gas accretion. 
In this way, it would be similar to Mrk\,1486, which \ct{cameron21} suggested has gone under a recent accretion event. However, only higher resolution \HI data may allow us to charachterise the full dynamical state of the gas disc and unveil its origin.

NGC\,4383 shows a biconical frustum geometry similar to most other galaxies. 
The estimated opening half-angle ($25^\circ-35^\circ$) is consistent with other galaxies hosting central starburst-driven outflows across a range of stellar masses and SFRs, such as Mrk\,1468, \cp[$\sim30^\circ$][]{mcpherson23}, M\,82 \cp[$\sim 25-30^\circ$][]{shopbell98,yuan23}  and NGC\,839 \cp[$<30^\circ$][]{rich10}. 

Considering the physical conditions within the outflow-dominated gas, NGC\,4383 shows a similar ionisation structure to M\,82 \cp{shopbell98} and NGC\,839 \cp{rich10}.
Namely, the base and the core of the outflow are consistent with being photoionised by recent, massive star formation, while emission line ratios get closer to the limit between photoionisation and ionisation by shocks towards the edges of the cone. 
Deriving the gas-phase metallicity from the strong emission lines, we find that the outflow is metal-enriched but not more so than the central star forming region or the disc. 
This metallicity trend is an interesting contrast to \ct{cameron21}, who find the outflow from Mrk\,1486 to be metal-enriched by $\sim0.2\,$dex compared to the disc average. 
We note the important caveat that we use strong line calibrations compared to the the $T_\mathrm{e}$-based method used by  \ct{cameron21}, which can lead to different absolute scalings.
Furthermore, most of the metals in outflows are carried by the hot ($T>10^6$\,K) gas phase. 
The ionised gas traced by these nebular lines in outflows arises from this hot phase mixing with the lower temperature and metallicity ISM entrained in the outflow. 
Thus, our metallicities are a lower limit on the true metallicity of the hot phase. 
However, we cannot say by how much, as this is strongly dependent on the complex interplay of mixing on the mass loading and launching conditions of outflows \cp[e.g.,][]{gronke18,fielding22}.

Our derived outflow velocities are broadly consistent with
expected values from literature scaling relations based on NGC\,4383's stellar mass ($\lgMstarMsun=9.44$) and integrated (UV+IR) star formation rate \cp[$\lgSFRMsunyr = 0.01$,][]{leroy19}. Equation set 14 from \ct{chisholm15} gives $v_\mathrm{out}$ values in the range $93-101\,\kms$, lower than the observed values, while the calibrations from \ct[][from their Figure 9]{xu22a} give expected outflow velocities of $v_\mathrm{out}=130-260\,\kms$, consistent with our observed values. \ct{heckman16} calibrate their scaling relations using $v_\mathrm{max}$, and for NGC\,4383 the expected $v_\mathrm{max} = 230-386\,\kms$, similarly consistent with our observed values.

Finally, our mass outflow rate and loading factor are within the ranges expected from literature scaling relations. 
However, these are definitely lower limits on the true mass outflow rate as the ionised gas phase does not trace a significant fraction of the outflowing mass. 
Instead, the colder, neutral gas phases entrained by the outflow carry most of the mass, for which \HI is a powerful tracer. 
Further work combining our MAUVE cubes with CO(2-1) data from the VERTICO survey and new MeerKAT \HI observations will enable us to obtain a more complete picture of NGC\,4383 
and better constrain the geometry and mass loading factors of its outflow \cp[e.g.,][]{yuan23}. 

There are clearly several challenges that remain in our understanding of outflow physics. 
Spatially resolving outflows with multiwavelength observations is key to make progress in this field, and this is becoming increasingly possible with MUSE, MeerKAT and ALMA as we increase our sample size of galaxies with outflows in the local Universe.

\section*{Acknowledgements}
We thank the referee, Timothy Heckman, for his comments that helped to improve the quality of this paper.

Based on observations collected at the European Southern Observatory under ESO programme(s): 105.208Y and 110.244E. 

We wish to acknowledge the custodians of the land on which this work was undertaken, the Wadjuk (Perth region) people of the Nyoongar nation, and their Elders past, present and future.
A.B.W. thanks Yifei Jin for helpful comments.
A.B.W. and L.C. acknowledge support from the Australian Research Council Discovery Project funding scheme (DP210100337).
L.C. is the recipient of an Australian Research Council Future Fellowship (FT180100066) funded by the Australian Government.
T.H.B. acknowledges support from the National Research Council of Canada via the Plaskett Fellowship of the Dominion Astrophysical Observatory.

Parts of this research were supported by the Australian Research Council Centre of Excellence for All Sky Astrophysics in 3 Dimensions (ASTRO 3D), through project number CE170100013.
The analysis in this work has been performed using the Python programming language, with the open-source package SciCM \url{https://github.com/MBravoS/scicm}.

We acknowledge support from AAO Data Central (\url{datacentral.org.au}).
\section*{Data Availability}
While the MAUVE data is currently proprietary until 2025, the collaboration is very open and readers interested in joining can visit the `team' page of the \href{https://mauve.icrar.org/}{\color{violet} MAUVE survey website}.



\bibliographystyle{mnras}
\bibliography{N4383} 



\appendix

\section{Assumed outflow geometry} \label{app:app1}
\begin{figure*}
    \centering
    \includegraphics[width=\textwidth]{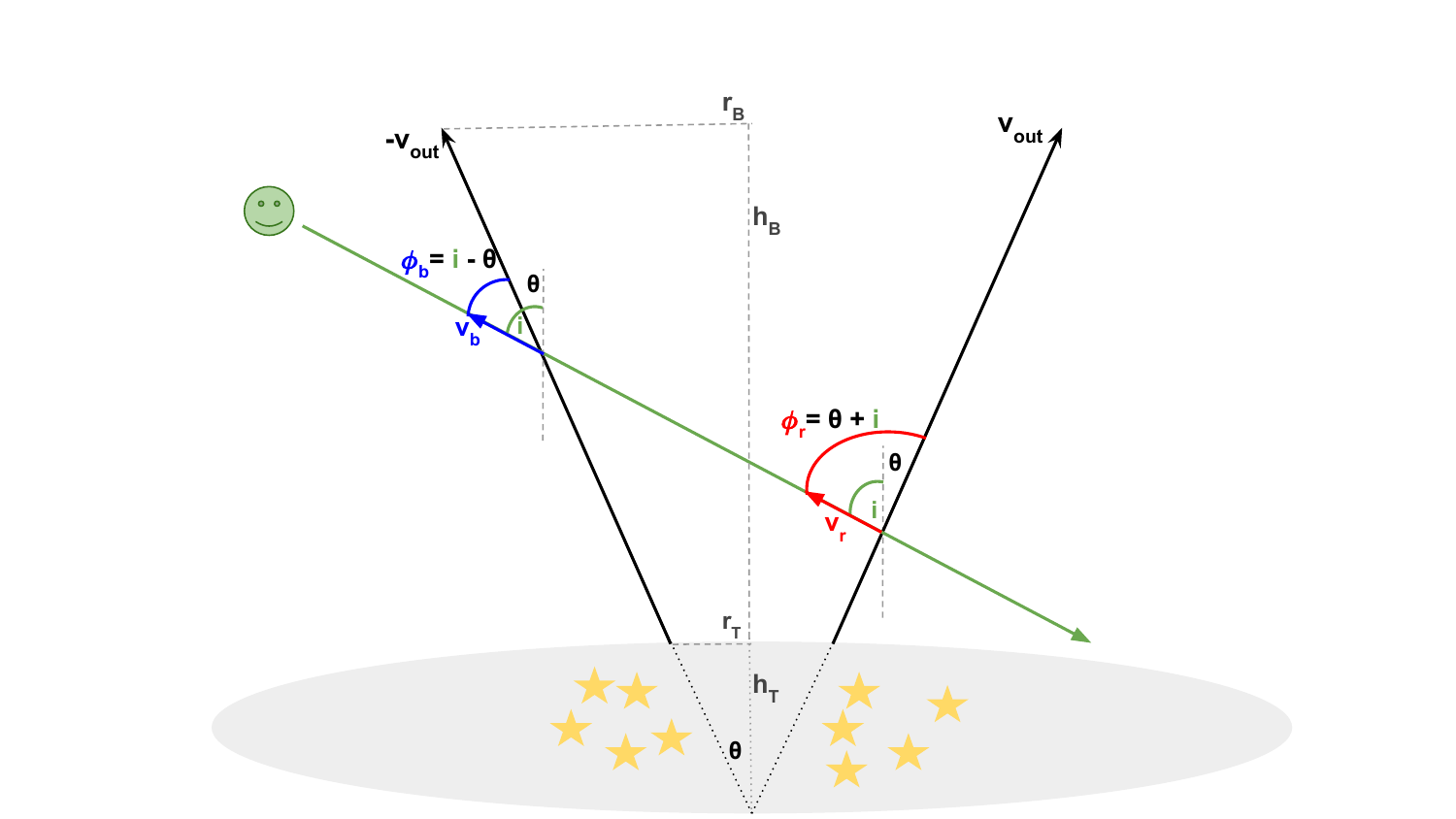}
    \caption{Adopted geometry for the NGC\,4383 outflow. Lower case subscripts (r and b) denote red- and blue-shifted velocity components. Upper case subscripts are used to refer to the spatial geometry, with `B' 
    and `T' indicating the base and the removed top of the frustum, respectively.}
    \label{fig:outflow_geo}
\end{figure*}
Here, we show the equations used when constraining the geometry of the outflow. 
In \fig{outflow_geo}, we show a diagram of our assumed geometry, a symmetric biconical frustum centered on the axis normal to the stellar disc plane with opening half-angle $\theta$, observed at an inclination of $i$. 
Using this geometry, the blueshifted and redshifted components of the velocity along the line of sight are 
\begin{align}
    v_\mathrm{b} &= -v_\mathrm{out}\cos \phi_\mathrm{b} \\
    v_\mathrm{r} &= v_\mathrm{out}\cos \phi_\mathrm{r}
\end{align}
where $\phi_\mathrm{b} = i-\theta$ and $\phi_\mathrm{r} = i+\theta$.
Rearranging eq A1 for $v_\mathrm{out}$, substituting into eq A2 and renaming $R=v_\mathrm{r}/(-v_\mathrm{b})$ we can write
\begin{align}
    R\cos \phi_\mathrm{b} &= \cos \phi_\mathrm{r} .
\end{align}
Expanding the cosine terms using angle summation/difference, collecting like terms, and simplifying
\begin{align}
    (1+R)\sin i \sin \theta &= (1-R)\cos i \cos \theta\\
    \tan i \tan \theta &= \frac{1-R}{1+R}.
\end{align}
Assuming an inclination for the galaxy enables us to solve for the opening half-angle:
\begin{align}
    \theta &= \arctan \Big(\frac{1-R}{1+R} \frac{1}{\tan i} \Big),
\end{align}
and $v_\mathrm{out}$ can then be calculated from eq A1 or eq A2.

It is useful to note that the relationship between radius and height for a cone also holds for the frustum, once the height of the excluded `top' (the smaller cone that would sit below the galaxy plane in this geometry) is accounted for
\begin{align}
    \tan\theta = \frac{r_T}{h_T} = \frac{r_B}{h_B+h_T}.
\end{align}
Knowing the radius where the outflow launches from the disc ($r_T$) lets us calculate the height of the removed `top'. 
Adding this to the (inclination corrected) observed maximum vertical extent of the outflow ($h_B$) lets us calculate the radius at this height $r_B$. 
The volume of the frustum is then given by 
\begin{equation}
    V = \frac{1}{3}\pi h_B (r_B^2 + r_B r_T + r_T^2)^3.
\end{equation}

\bsp	
\label{lastpage}
\end{document}